\documentclass[a4paper,11pt,final]{article}

\usepackage{authblk}

\usepackage{graphics}
\usepackage{graphicx}
\usepackage{epsfig}

\usepackage{amssymb}
\usepackage{amsthm}

\usepackage{amsmath}
\date{}

\begin{document}


\title{On the influence of reflective boundary conditions on the
statistics of Poisson-Kac diffusion processes}


\author[1]{Massimiliano Giona$^*$}
\author[2]{Antonio Brasiello}
\author[3]{Silvestro Crescitelli}
\affil[1]{Dipartimento di Ingegneria Chimica DICMA
Facolt\`{a} di Ingegneria, La Sapienza Universit\`{a} di Roma 
via Eudossiana 18, 00184, Roma, Italy  \authorcr
$^*$  massimiliano.giona@uniroma1.it}                                          

\affil[2]{Dipartimento di Ingegneria Industriale
Universit\`{a} degli Studi di Salerno
via Giovanni Paolo II 132, 84084 Fisciano (SA), Italy}

\affil[3]{Dipartimento di Ingegneria Chimica,
 dei Materiali e della Produzione Industriale
Universit\`{a} degli Studi di Napoli ``Federico II''
piazzale Tecchio 80, 80125 Napoli, Italy}
\maketitle

\begin{abstract}
We analyze the influence of reflective boundary conditions
on the statistics of Poisson-Kac diffusion processes, and
specifically how they modify the Poissonian switching-time statistics.
After addressing simple cases such as diffusion in a channel,
and the switching statistics in the presence of a polarization
potential, we thoroughly study Poisson-Kac diffusion
in fractal domains. Diffusion in fractal spaces
highlights neatly how the modification in the switching-time
statistics associated with  reflections against a complex and fractal
boundary induces new emergent features of Poisson-Kac  diffusion
leading to a transition from a regular behavior at shorter timescales
to emerging anomalous diffusion properties controlled
by walk dimensionality of the fractal set. 
\end{abstract}

\section{Introduction}
\label{sec_1}

The first, and certainly most intuitive microscopic description 
of diffusion processes is certainly via particle Brownian  motion,
dating to a series of papers by   A. Einstein \cite{einstein} and 
J. Perrin \cite{perrin}.
Mathematically, the connection between
 this archetype of irreversibility (diffusion),
and random motion at particle level  can be formulated in
a rigorous way, starting from the definition of stochastic
integration (Ito, Stratonovich, etc.) as a particular form
of Stieltjes integration, introducing the concept of
stochastic differential equations (Langevin equations),
and deriving their statistical properties, thus finally
obtaining a parabolic equation for the
probability density function (the forward Fokker-Planck equation)
\cite{gardiner,risken,ottinger}.
The mathematical tool establishing the connection between stochastic dynamics
at microscales, and macroscopic diffusion processes is the
concept of Wiener processes (also referred to as mathematical Brownian
motion). A one-dimensional Wiener process ${\mathcal W}(t)$
is a continuous stochastic process possessing independent increments
that are distributed in a Gaussian way (with zero mean and square
variance equal to the time elapsed).

Therefore, a diffusion equation of the form
\begin{equation}
\partial_t p(x,t) = D \, \partial_x^2 p(x,t)
\label{eq1_1}
\end{equation} 
can be viewed as the Eulerian description of
a microscopic Lagrangian motion described by
the stochastic kinematics
\begin{equation}
d x(t)= \sqrt{2 \, D} \, d w(t) \; ,
\label{eq1_2}
\end{equation}
where $d w(t)$ are the infinitesimal increments
of a one-dimensional Wiener process in the time interval
$(t,t+dt)$.

Both eqs. (\ref{eq1_1}) and ({\ref{eq1_2}) possess fundamental
physical limitations. The diffusion equation (\ref{eq1_1})
is characterized by a unbounded propagation velocity that
manifestly violates the basic principles of the theory of
special relativity. This
property is one-to-one with the
fractal nature of the graph of a realization $x(t)$ vs $t$
of the stochastic differential equation (\ref{eq1_2}),
which, as well known, possesses fractal dimension $3/2$ \cite{falconer}.

In order to solve the above mentioned problem, stochastic
models possessing finite propagation velocity have
been proposed, essentially based on the use of Poisson
processes for generating the stochastic perturbations
(see Section \ref{sec_2}) \cite{goldstein,kac,rosenau,kolesnik1,kolesnik2}.
In the simplest case of one-dimensional spatial problems,
the particle moves with constant and bounded velocity,
and 
the stochastic perturbation acts simply by  changing the
velocity direction (velocity switching), and the 
statistics of the switching times  follows 
 an exponential distribution.
The kinematic equations associated with this
simple random motion are relativistically consistent and describe a 
family of stochastic processes that almost everywhere (i.e. apart from
 the
time instants when the switching of the velocity direction
 occurs) differentiable functions
of time.

The basic article addressing this simple stochastic model is due to
  M. Kac
published in 1974 \cite{kac}, in connection with a stochastic interpretation
of the telegraphers' equation. For the sake of historical
correctness, the first paper dealing with this class
of models is by S. Goldstein \cite{goldstein} that considered similar
processes referred to as persistent random walk.
Subsequently, Gaveau et al. applied this model to
derive, via analytic continuation, the one-dimensional Dirac
equation of relativistic quantum mechanics \cite{kac_dirac}.

A significant and fertile Literature  originates from these
observations, using Poisson processes as a model
of bounded noise, or as a prototype of colored noise (due to
the exponential decay of the correlation function), see \cite{dicho1,dicho2,dicho3,dicho4,dichox,dicho5,dicho6} just to quote some of the Literature
 covering different aspects of physical interest.

Henceforth, we will refer to this class of processes as
Poisson-Kac processes (see Section \ref{sec_2} for details).
In the physical Literature, Poisson-Kac processes are
often referred to as dichotomous noise.

While at smaller timescales (smaller than
the average time interval between two consecutive velocity switching) the
motion is smooth, at larger timescales Poisson-Kac processes
retain emergent Brownian properties and, under certain assumptions
on the value of the model parameters, they converge to
the ordinary Brownian motion case (Kac limit) \cite{kac}.

In a previous article \cite{brasiello}, we have studied the limitations imposed
by the wave-like nature of the Poisson-Kac processes on the
representation of physically plausible boundary conditions
to be applied to the (Eulerian) balance equations  for
the probability density function (actually, for the
probability partial waves, see further Section \ref{sec_2}).

In this article, we analyze the impact of reflective boundary
conditions, i.e. of impearmability conditions
in a closed system, on the
statistics of the switching times of Poisson-Kac diffusion
processes.

Indeed, due to the wave-like nature of these processes,
any form of reflection modifies the exponential
switching-time statistics. Reflection conditions act as 
an active modulation of the Poissonian statistics. Specifically,
complex boundary conditions are able to modify the long-term
properties of the Poisson-Kac diffusion processes determining
anomalous diffusive scalings. This means that the long-term (emerging)
fractal properties of the Poisson-Kac processes
are altered by complex  geometry of the boundary.

The article is organized as  follows. Section \ref{sec_2} provides  
a brief description of Poisson-Kac processes functional to
the further  developments of the present article, 
by considering also two-dimensional models
that are further analyzed in Section \ref{sec_5}. Section \ref{sec_3} 
introduces the issue
of the influence of reflection boundary conditions on the
Poissonian statistics  in a simple case, namely that of particles
in a straight two-dimensional channel (a form of Knudsen diffusion problem \cite{knudsen}).
Section \ref{sec_4} addresses another relatively simple, albeit interesting
model, namely diffusion in the presence of a potential determining
polarization effects at the walls \cite{brasiello}.
Section \ref{sec_5} addresses the case of fractal boundary conditions
by considering Poisson-Kac diffusion on fractal sets 
(treated in a continuum way,
and not merely as a lattice diffusion problem) \cite{havlin,havlin1}. 
 This is the first application of Poisson-Kac processes 
in disordered systems. This model can be also viewed as a stochastic microscopic
description of a linear viscoelastic transport problem (transport
problem with memory) in a fractal
medium. However, the main focus in this article is 
on reflection conditions and how they   
modify the emerging long-term fractal properties of a Poisson-Kac
diffusive trajectory.

\section{Statement of the problem}
\label{sec_2}

Consider the  one-dimensional   problem of a particle moving
in a deterministic potential  $U(x)$ under overdamped conditions.
If $\eta$ is the friction factor, the particle experiences a deterministic
velocity field $v(x)$ given by
\begin{equation}
v(x)= - \frac{1}{\eta} \, \partial_x U(x) \, ,
\label{eq2_1}
\end{equation}
to which stochastic fluctuations are superimposed.
Therefore, the equation of motion can be written as
\begin{equation}
d x(t) = v(x(t)) \, d t + \frac{1}{\eta} \, d F_{\rm stoca}(t) \, ,
\label{eq2_2}
\end{equation}
where $F_{\rm stoca}(t)$ is the stochastic force, and $d F_{\rm stoca}(t)$
its infinitesimal increment.
In the case of the classical overdamped Langevin equation,
$F_{\rm stoca}(t)= \eta \, \sqrt{2 \, D_0} \, w(t)$ where $w(t)$ is
a Wiener process \cite{ottinger}. In the case of the finite propagation 
velocity model
proposed by Kac, $F_{\rm stoca}(t)$ is the integral over time
\begin{equation}
F_{\rm stoca}(t)= b \, \eta \, \int_0^{t} (-1)^{\chi(\tau)} \, d \tau \, ,
\label{eq2_3}
\end{equation}
where $\chi(t)$ is a  Poisson process possessing  the switching rate
$a>0$, i.e., $E[\chi(t)/\chi(0)=0]= a t$, and $b>0$.

The Poisson-Kac equation corresponding to eq. (\ref{eq2_4}) in
the presence of a deterministic velocity field $v(x)$ is given by
\begin{equation}
d x(t) = v(x(t)) \, d t + b \, (-1)^{\chi(t)} \, d t \; .
\label{eq2_4}
\end{equation}
Let us indicate with  $X(t)$ the stochastic process associated
with the equation (\ref{eq2_4}) at time $t$.

Since $(-1)^{\chi(t)}$ can attain solely two values $\pm 1$, the
 statistical description of the process is based on the 
two partial probability density functions $p^\pm(x,t)$,
\begin{equation}
p^\pm(x,t) dx = \mbox{Prob} \{ X(t) \in (x,x+dx), \; (-1)^{\chi(t)} = \pm 1 \} \; ,
\label{eq2_5}
\end{equation}
satisfying the 
wave equations with recombination (dissipative
wave equations)
\begin{eqnarray}
\partial_t p^+(x,t) &  = & - \partial_x \left [ (v(x)+b)  \, p^+(x,t)\right ]
-a \, p^+(x,t)+ a \, p^-(x,t) \nonumber \\
\partial_t p^-(x,t) & = & - \partial_x \left [ (v(x)-b) \, p^-(x,t)\right ]
+a \, p^+(x,t)- a \, p^-(x,t) \, .
\label{eq2_9}
\end{eqnarray}
As these equations correspond to forward and backward 
 propagating waves, the
partial probabilities $p^\pm(x,t)$ are also referred to
as {\em partial probability waves}.

The probability density function $p(x,t)$ for $X(t)$ at
time $t$ is the sum of the two partial probability
densities $p^\pm(x,t)$ 
\begin{equation}
p(x,t)=p^+(x,t)+p^-(x,t) \; .
\label{eq2_8}
\end{equation}

Let us  briefly address  the properties of the stochastic 
trajectories associated with eq. (\ref{eq2_4}), and the inclusion
of impermeability conditions in finite domains.

First, consider eqs. (\ref{eq2_4}) and (\ref{eq2_9}) in the free space i.e.
as the unconstrained  propagation along the real line.
As $\chi(t)$ is distributed in a Poissonian way, the statistics
of the switching times $\tau$, i.e. thta of the time intervals
between two consecutive switchings of $(-1)^{\chi(t)}$
is distributed in an exponential way, according to
the probability density functions $g_P(\tau)$
\begin{equation}
g_P(\tau) = a \, e^{-a \tau} \; .
\label{eq2_10}
\end{equation}
This is the {\em bare} Poissonian result.
The exponential probability density function
 for the switchnig times gives rise
to the linear recombination terms $\pm (-a p^+ + a p^-)$
entering the balance equations for the partial probabilities
(\ref{eq2_9}), determining the  exchange from $p^+$ to $p^-$
and viceversa  per unit time at a constant rate equal to $a$.

 Correspondingly, if $a$ and $b$ are
bounded, the graph of a generic realization of eq. (\ref{eq2_4})
as a function of time $t$ is an  almost everywhere  smooth function in all
the  open intervals between two consecutive
switchings. If $t=t^*$ is a time instant at which a switching event
occurs,  solely the derivative of $x(t)$ is
discontinuous at $t=t^*$, still keeping a bounded left and
right derivative.

Let us suppose that eq. (\ref{eq2_4}) describes particle
motion in a box, and that $x=0$ and $x=L>0$ are the positions
of the box walls, so that particles moving in $(0,L)$
cannot cross them or perform any sort of tunneling at them.
Suppose for simplicity that the deterministic contribution $v(x)$
is absent.
 Impermeability at $x=0,L$
implies total reflections for the partial waves $p^\pm(x,t)$.
This means  that at $x=0$  the backward propagating waves $p^-(x,t)$ is
totally reflected, and this determines the boundary condition for $p^+(x,t)$,
i.e.,
\begin{equation}
p^+(0,t)= p^-(0,t) \, .
\label{eq2_11}
\end{equation}
In a similar way, the forward wave $p^+(x,t)$ is totally
reflected at $x=L$, and this induces the complementary boundary
condition
\begin{equation}
p^-(L,t)=p^+(L,t) \, .
\label{eq2_12}
\end{equation}
In terms of stochastic trajectories, reflection at an impermeable
wall corresponds to an additional switching of $(-1)^{\chi(t)}$,
externally induced by the boundaries,  that superimposes
to the Poissonian exponential statistics. If this occurs at $x=0$,
this means that the wall forces the transition of $(-1)^{\chi(t)}$
from $-1$ to $1$, and similarly at $x=L$, the wall-imposed switching
from $1$ to $-1$ occurs.

Consequently, the presence of impermeable boundary conditions provides
an active contribution  to the statistical properties
of the trajectories of the stochastic process (\ref{eq2_4}) determining
a modification of the effective switching time statistics
$g(\tau)$, which is no longer equal to the bare distribution
$g_P(\tau)$ expressed by eq. (\ref{eq2_10}).
The analysis of $g(\tau)$ and of its consequences on the emerging
properties of the solutions of eq. (\ref{eq2_4}) are the main
issues of this article.

To conclude this brief introduction to Poisson-Kac processes,
let us mention the connection between eq. (\ref{eq2_4}) and the
classical Langevin equation driven by Wiener fluctuations.
For $a,b \rightarrow \infty$, keeping fixed the ratio
\begin{equation}
\frac{b^2}{2 a} = D \, ,
\label{eq2_13}
\end{equation}
the Poisson-Kac stochastic equation (\ref{eq2_4}) converges
to the classical Langevin equation
\begin{equation}
d x(t)= v(x(t)) \, dt + \sqrt{2 D} \, d w(t) \, ,
\label{eq2_14}
\end{equation}
and the probability density function $p(x,t)$ converges to
the solution of the Fokker-Planck equation (for mathematical details
see also \cite{janssen}) 
\begin{equation}
\partial_t p(x,t) = - \partial_x \left [ v(x) \, p(x,t) 
\right ] = D \, \partial_x^2 p(x,t) \; .
\label{eq2_15}
\end{equation}
This property is referred to as the {\em Kac limit}.

\subsection{Two-dimensional Poisson-Kac processes}
\label{sec_2_1}

Since in Section \ref{sec_5} we   consider Poisson-Kac processes
in complex two-dimensional structures (fractal sets), it is
useful to address here the extension of eq. (\ref{eq2_4}) in higher
dimensions, which is an issue scarsely addressed in the physical
Literature \cite{masoliver_multi,plyukhin}, but very fertile
in theoretical probability theory \cite{kolesnik1,kolesnik2}.

Although there are very interesting alternative extensions
of Poisson-Kac processes in higher dimensions,
mainly addressed by Kolesnik and coworkers \cite{kolesnik1,kolesnik2},
 we focus here  on
the most natural and straightforward generalization of eq. (\ref{eq2_4})
in two-dimensional domains and in the absence
of external potentials or deterministic velocity
contributions.

The simplest generalization of eq. (\ref{eq2_4}) in two-dimensional
spatial coordinates in the case of pure stochastic motion 
considers two independent Poisson processes $\chi_1(t)$,
$\chi_2(t)$, each of which acts on distinct spatial coordinates,
i.e.,
\begin{eqnarray}
d x(t) & = & b \, (-1)^{\chi_1(t)} \, dt  \nonumber \\
d y(t) & = & b \, (-1)^{\chi_2(t)} \, dt
\label{eq2_16}
\end{eqnarray}
possessing the same statistical properties, i.e.,
\begin{equation}
E[\chi_1(t) /\chi_1(0)=0]=
E[\chi_2(t)/\chi_2(0)=0]=a t \; .
\label{eq2_16a}
\end{equation}
Analogously to the one-dimensional case, the statistical
properties of the purely diffusive Poisson-Kac model
(\ref{eq2_16}) are fully described by the
system of four partial probability density function
$p^{(\pm,\pm)}({\bf x},t)$, here ${\bf x}=(x,y)$,
\begin{equation}
p^{(\pm,\pm)}({\bf x},t) \, d {\bf x} =
\mbox{Prob} \left \{ {\bf X}(t) \in ({\bf x},{\bf x}+d {\bf x}) , \;
(-1)^{\chi_1(t)} = \pm 1, \; (-1)^{\chi_2(t)} = \pm 1 \right \} \, ,
\label{eq2_16bis}
\end{equation}
where $d {\bf x}$ indicates the two-dimensional area element,
and ${\bf X}(t) =(X(t),Y(t))$ is the two-dimensional stochastic
process described by eq. (\ref{eq2_16}) at time $t$.

For the sake of notational simplicity, let us 
use the notation $p_1({\bf x},t)=p^{(+,+)}({\bf x},t)$,
$p_2({\bf x},t)=p^{(-,+)}({\bf x},t)$, $p_3({\bf x},t)=p^{(-,-)}({\bf x},t)$,
$p_4({\bf x},t)=p^{(+,-)}({\bf x},t)$, so that the overall probability
density function for ${\bf X}(t))$ is the sum of the $p_h$'s,
\begin{equation}
p({\bf x},t)= \sum_{h=1}^4 p_h({\bf x},t) \, .
\label{eq2_17}
\end{equation}
It can be easily recognized that each $p_h({\bf x},t)$  
describes the statistics of a particle ensemble moving with constant
velocity $b \, \boldsymbol{\beta}_h$, where
\begin{equation}
\boldsymbol{\beta}_h = \left ( \cos \left ( \frac{ 2 (h-1) \pi + \pi}{4}
\right ) , \sin \left ( \frac{ 2 (h-1) \pi + \pi}{4}
\right ) \right ) \, ,
\label{eq2_18}
\end{equation}
$h=1,\dots,4$. As it regards the recombination dynamics, consider for
example $p_1=p^{(+,+)}$, and use the notation
$(\pm,\pm)$ to indicate  $\left ( (-1)^{\chi_1(t)}= \pm 1, (-1)^{\chi_2(t)}= \pm 1  \right )$. Considering the statistics
of an ensemble of particles, the  switching in the unit time
 from $(+,+)$ to
$(-,+)$ and $(+,-)$ occurs with rate $a$
 The reverse processes possess
obviously the same rate $a$.
Consequently, the balance equation for $p_1({\bf x},t)$ is
the Eulerian continuity equation for a particle ensemble moving
with  constant velocity $b \,  \boldsymbol{\beta}_1$ which accounts
for the above discussed recombination dynamics
amongs partial waves,
\begin{equation}
\partial_t  \, p_1({\bf x},t) = - b \, \nabla \cdot \left ( \boldsymbol{\beta}_1
\, p_1({\bf x},t)  \right ) - 2 a \, p_1({\bf x},t) + a \left ( p_2({\bf x},t)+p_4({\bf x},t)
\right ) \, .
\label{eq2_19}
\end{equation}
where $\nabla$ indicates the nabla operator with respect to the spatial coordinates ${\bf x}$.
In a similar way, the balance equations for the other partial waves follow
\begin{eqnarray}
\partial_t \, p_2({\bf x},t)   =  - b \, \nabla \cdot \left ( \boldsymbol{\beta}_2
\, p_2({\bf x},t)  \right ) - 2 a \, p_2({\bf x},t) + a \left ( p_1({\bf x},t)+p_3({\bf x},t) \right )  \nonumber \\
\partial_t  \, p_3({\bf x},t)   =  - b \, \nabla \cdot \left ( \boldsymbol{\beta}_3
\, p_3({\bf x},t)  \right ) - 2 a \, p_3({\bf x},t) + a \left ( p_2({\bf x},t)+p_4({\bf x},t) \right )
\label{eq2_20} \\
\partial_t \, p_4({\bf x},t)   =  - b \, \nabla \cdot \left ( \boldsymbol{\beta}_4
\, p_4({\bf x},t)  \right ) - 2 a \, p_4({\bf x},t) + a \left ( p_1({\bf x},t)+p_3({\bf x},t) \right )  \, . \nonumber
\end{eqnarray}
Gathering eqs. (\ref{eq2_19})-(\ref{eq2_20}) and summing over the index ``$h$''
 of the
partial waves, the balance equation for the 
overall probability density function follows
\begin{equation}
\partial_t \, p({\bf x},t) = - \nabla \cdot {\bf J}({\bf x},t) \, ,
\label{eq2_21}
\end{equation}
where ${\bf J}({\bf x},t)$ is the probability flux originating
from the stochastic bivariate Poissonian perturbation
\begin{equation}
{\bf J}({\bf x},t)= b \, \sum_{h=1}^4 \boldsymbol{\beta}_h \, p_h({\bf x},t)
\, .
\label{eq2_22}
\end{equation}
From eqs.  (\ref{eq2_19})-(\ref{eq2_22}), it is straightforward to obtain
the constitutive equation for the probability flux, multiplying each
evolution equation for $p_h({\bf x},t)$
by $b \, \boldsymbol{\beta}_h$ and summing over the index ``$h$'',
\begin{equation}
\partial_t \, {\bf J}({\bf x},t) = - b^2 \, \nabla \cdot \left (
\sum_{h=1}^4 \boldsymbol{\beta}_h \boldsymbol{\beta}_h \, p_h({\bf x},t)
\right ) - 2 a \, {\bf J}({\bf x},t) \, .
\label{eq2_23}
\end{equation}
Several observations follows from the system of equations
(\ref{eq2_21})-(\ref{eq2_23}): 
\begin{itemize}
\item the statistical description of the process corresponds
to that of a linear viscoelastic material with memory, as
the constitutive equation for ${\bf J}({\bf x},t)$ depends on
the first-order time derivative of the flux itself;
\item this representation is however {\em irreducible} in terms of
$p({\bf x},t)$ and ${\bf J}({\bf x},t)$, as the correct description
of the process involves the full system of partial probability  waves
$\{p_h({\bf x},t) \}_{h=1}^4$;
\item although there are some analogy between eq. (\ref{eq2_23})
and the Cattaneo transport equation with memory in higher dimensions
\cite{cattaneo,jou},
the functional form of the constitutive equation (\ref{eq2_23})
marks the fundamental difference between the two models.
For finite values of $a$ and $b$, it is impossible to
express the time-derivative of the probability flux ${\bf J}({\bf x},t)$ solely
as a function of ${\bf J}({\bf x},t)$ itself and of the overall
probability density function $p({\bf x},t)$: all the systems of $\{ p_h({\bf x},t) \}_{h=1}^4$ should be considered.
If the first term at the r.h.s. of eq. (\ref{eq2_23}) is approximated
by a term proportional to the gradient of $p({\bf x},t)$, as in the
two-dimensional Cattaneo model \cite{cattaneo,jou}, the
stochastic interpretation of the resulting constitutive equation is 
completely lost, and this is the fundamental reason while the higher-dimensional
Cattaneo model does not satisfy the requirement of positivity \cite{brasiello};
\item the classical diffusion equation follows 
from (\ref{eq2_21})-(\ref{eq2_23}) in the Kac limit.
\end{itemize}

To prove the last observation, consider the limit for
arbitrarily large $a$ and $b$ ($a,b \rightarrow \infty$),
keeping fixed the ratio $b^2/a$ to a constant value.
If the switching rate tends to infinity, the recombination
process between partial waves becomes infinitely fast, and
the structure of the partial waves ``thermalizes'' at any ${\bf x}$,
meaning that all the $p_h({\bf x},t)$ collapse into a unique
function that is just $p({\bf x},t)/4$.
Enforcing this property into  eq. (\ref{eq2_23}) the constitutive
equation 
for the probability flux becomes
\begin{equation}
{\bf J}({\bf x},t) = - \frac{b^2}{2 a} \, \nabla \cdot \left ( {\bf B} \, p
\right ) - \frac{1}{2 a} \, \partial_t \, {\bf J}({\bf x},t) \, ,
\label{eq2_24}
\end{equation}
where ${\bf B}$ is the diadic tensor
\begin{equation}
{\bf B}= \frac{1}{4} \, \sum_{h=1}^4 \boldsymbol{\beta}_h \boldsymbol{\beta}_h
\, .
\label{eq2_25}
\end{equation}
Because of the structure of the normalized velocities
$\boldsymbol{\beta}_h$ defined by eq. (\ref{eq2_18}), this tensor is isotropic
and equal to the identity tensor, ${\bf B}={\bf I}$.
In the Kac limit, the second term at the right hand side of eq. (\ref{eq2_24})
vanishes and the constitutive equation reduces to
the Fickian form, ${\bf J}({\bf x},t) = - D \, \nabla p({\bf x},t)$,
where the diffusivity is given by eq. (\ref{eq2_13}).

Finally,  let us discuss the role of impermeability conditions.
Consider for simplicity the problem of Poisson-Kac diffusion
inside a square, $(x, y) \in   (0,L)\times (0,L)$. Because
of the symmetries of the normalized
velocities $\boldsymbol{\beta}_h$, each reflection
at the box walls, placed at $x=0,L$, $y=0,L$ corresponds
to an externally forced switching either of $(-1)^{\chi_1(t)}$
(for the reflections at $x=0,L$) or of $(-1)^{\chi_2(t)}$
(for the reflections at $y=0,L$). The boundary conditions for
the partial waves at the walls can be obtained
either by enforcing that the normal component of the probability flux 
should vanish at the box walls or by expressing total reflection
in terms of the partial waves as it result from the switching
of one of the $(-1)^{\chi_h(t)}$, $h=1,2$. For example, consider
the collisions with the boundary at $x=0$. In this case
$(-1)^{\chi_1(t)}$ switches from the value $(-1)^{\chi_1(t_-)} =-1$ to 
the value $(-1)^{\chi_1(t_+)}=1$, and consequently
the boundary conditions become
\begin{equation}
p_1(0,y,t)=p_3(0,y,t) \, , \qquad  p_4(0,y,t)=p_2(0,y,t) \, ,
\label{eq2_26}
\end{equation}
meaning that the two backward waves (along the $x$ coordinate) $p_3$, and $p_2$
are reflected back into $p_1$ and $p_4$, respectively.
Analogous expressions can be derived for the other wall collisions.

\section{The simplest case: Knudsen  effect}
\label{sec_3}

The simplest physical situation where the effects of impermeable walls
controls the statistics of Poisson-Kac
diffusion processes is represented by particle motion in straight
channels under creeping conditions, so that the deterministic velocity
fields acts only along the axial direction. Consider a two-dimensional model
letting $x$ be the axial, and $y$ the transverse coordinates.

Particle motion is described (using nondimensional variables)
by the system of stochastic equations
\begin{eqnarray}
d x (t) & = & v(y(t)) \, dt + b \, (-1)^{\chi_1(t)} \, dt \nonumber \\
d y(t)  & =  & b \, (-1)^{\chi_2(t)} \, d t
\label{eq3_1}
\end{eqnarray}
where $y \in (0,1)$, equipped with reflection boundary conditions 
at $y=0,1$ which represent channel walls. As the transverse motion
is completely decoupled from the axial one, we can 
consider the second equation (\ref{eq3_1}) as a stand-alone
problem. The Poisson process $\chi_2(t)$ is characterized
by a switching rate $a$, and we assume  $b$ and $a$ to be
related by the Kac condition (\ref{eq2_13}), i.e. $b^2 = 2 D a$,
where $D$ is the nondimensional diffusion coefficient.
Therefore, instead of $a$ and $b$ we can use $b$ and $D$ or $a$ and $D$
to define completely the characteristic of the stochastic perturbation
that controls transverse particle motion.
ppp
Figure \ref{Fig0} depicts the time series of the consecutive switching intervals $\tau_n$ vs the numeral order $n$. The most evident effect of
impermeable  boundaries is that the switching intervals $\tau_n$ are upper bounded,
as the maximum attainable value of the switching time is given by
$\tau_{\rm max}=L/b$, where $L$ is the transverse width of the channel,
and $b$ the characteristic velocity of the stochastic
perturbation. In the present case, $L=1$, and $b=\sqrt{2 \, D \, a}$, so that
\begin{equation}
\tau_{\rm max}= \frac{1}{\sqrt{2 \, D \, a}} \, .
\label{eq3_2}
\end{equation}
For $D=a=1$, $\tau_{\rm max}=1/\sqrt{2}$, corresponding to the
horizontal dashed line in figure \ref{Fig0}.

\begin{figure}[h!]
\begin{center}
\epsfxsize=11cm
\epsffile{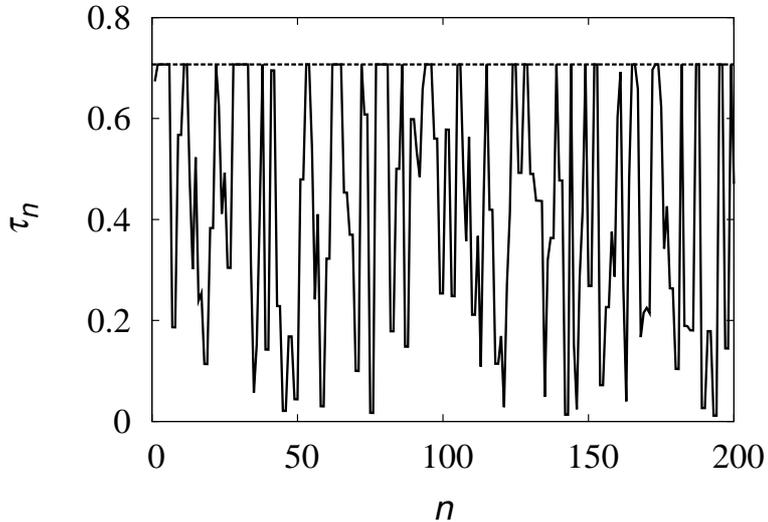}
\end{center}
\caption{Time series of the switching intervals $\tau_n$ vs $n$
for a  Poisson-Kac diffusive trajectory at $D=1$, $a=1$.}
\label{Fig0}
\end{figure}

This result follows from the elementary observation that the
longest switching time is reached by a particle hitting  consecutively
the two walls without switching the velocity direction due to
the Poissonian statistics. Consequently, the probability
density function $g(\tau)$ for the switching times admits a compact support
represented by the interval $[0,\tau_{\rm max}]$, while the
bare Poissonian $g_P(\tau)$ does  not.

This effect can be referred to as the {\em Knudsen effect} on 
the Poissonian statistics induced by wall reflection, to mark
the analogy with the phenomenon of Knudsen diffusion
for dilute gases in narrow pores where the collision
with the pore walls dominate the diffusive transport process.

Figure \ref{Fig1} depicts the behavior of the switching time
pdf $g(\tau)$ obtained  from the stochastic simulation of eq. (\ref{eq3_1}).
The  pdf  $g(\tau)$ can be approximated by the superposition of a 
smooth exponential distribution defined in $[0,\tau_{\rm max}]$ possessing
a decay exponent $a_e \neq a$, different from the
Poissonian switching rate $a$,  and of an impulsive contribution
centered at $\tau_{\rm max}$, i.e.,
\begin{equation}
g(\tau) = A \, e^{- a_e \tau} + B \, \delta(\tau-\tau_{\rm max}) \, ,
\label{eq3_3}
\end{equation}
where $A$ and $B$ are positive constants, such that
$A (1-e^{-a_e \tau_{\rm max}})/a_e + B=1$.

\begin{figure}[h!]
\begin{center}
\epsfxsize=11cm
\epsffile{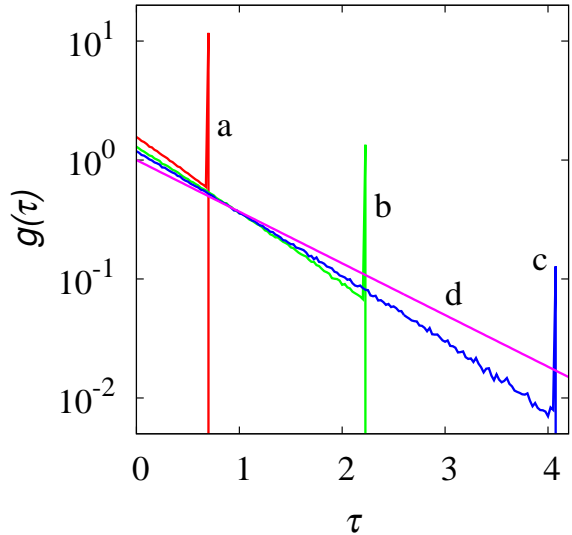}
\end{center}
\caption{Switching time probability density function $g(\tau)$
vs $\tau$ for the one-dimensional
 Kac diffusion process on the unit interval for
$a=1$ at different values of the diffusivity $D$.
Line (a):  $D=1$, line (b): $D=0.1$, line (c) $D=0.03$. Line
(d) corresponds to the pure Poisson statistics $g_{P}(\tau)=ae^{-a \, \tau}$
in the absence of boundary condition effects.}
\label{Fig1}
\end{figure}

The effective rate depends on $a$ and $D$, and its rescaled
behavior, i.e., $(a_e-a)/a$,
 is depicted in figure \ref{Fig1a} as a function of $a$
for different values of $D$. As can be observed, the relative
deviation from the Poissonian rate $a$ is more pronounced
as $D$ and $a$ increase.

\begin{figure}[h!]
\begin{center}
\epsfxsize=11cm
\epsffile{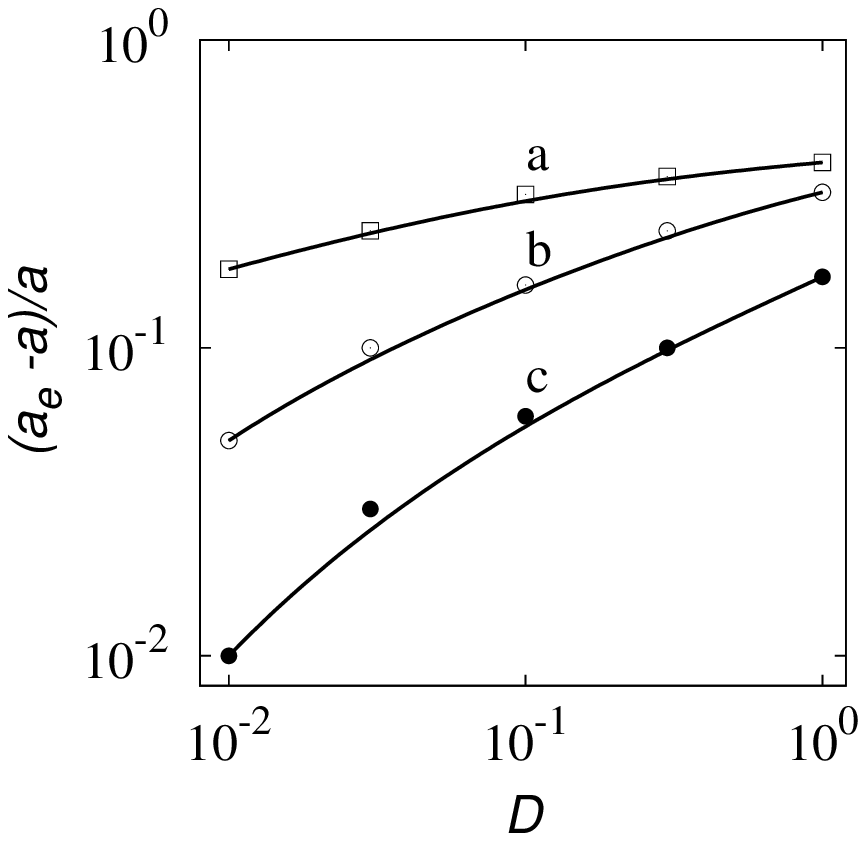}
\end{center}
\caption{Rescaled effective switching rate $(a_{e}-a)/a$ vs diffusivity $D$ at 
three different values of $a$. Line (a) and ($\square$): $a=1$,
line (b) and ($\circ$): $a=0.1$, line (c) and ($\bullet$): $a=0.01$.}
\label{Fig1a}
\end{figure}

It is possible to obtain a quantification of the distortion
effects induced by wall reflections using simple arguments.
Consider the fraction $\phi_{\rm wall}$ of the number of switching
events associated with wall collisions with respect to the
overall number of switchings. A lower bound for $\phi_{\rm wall}$
can be obtained by considering that this fraction should 
necessarily be greater than the Poissonian probability 
of having switching times greater than $\tau_{\rm max}$, i.e.,
\begin{equation}
\phi_{\rm wall} \geq \int_{\tau_{\rm max}}^\infty g_P(\tau) \, d \tau
= e^{-a \tau_{\rm max}} =e^{- \sqrt{a/2 D}} = \phi_0(a,D) \, .
\label{eq3_4}
\end{equation}
Figure \ref{Fig2} depicts the behavior of $\phi_{\rm wall}$
and of its lower-bound estimate $\phi_0(a,D)$ as a function 
of $a$ for two values of $D$.
\begin{figure}[h!]
\begin{center}
\epsfxsize=11cm
\epsffile{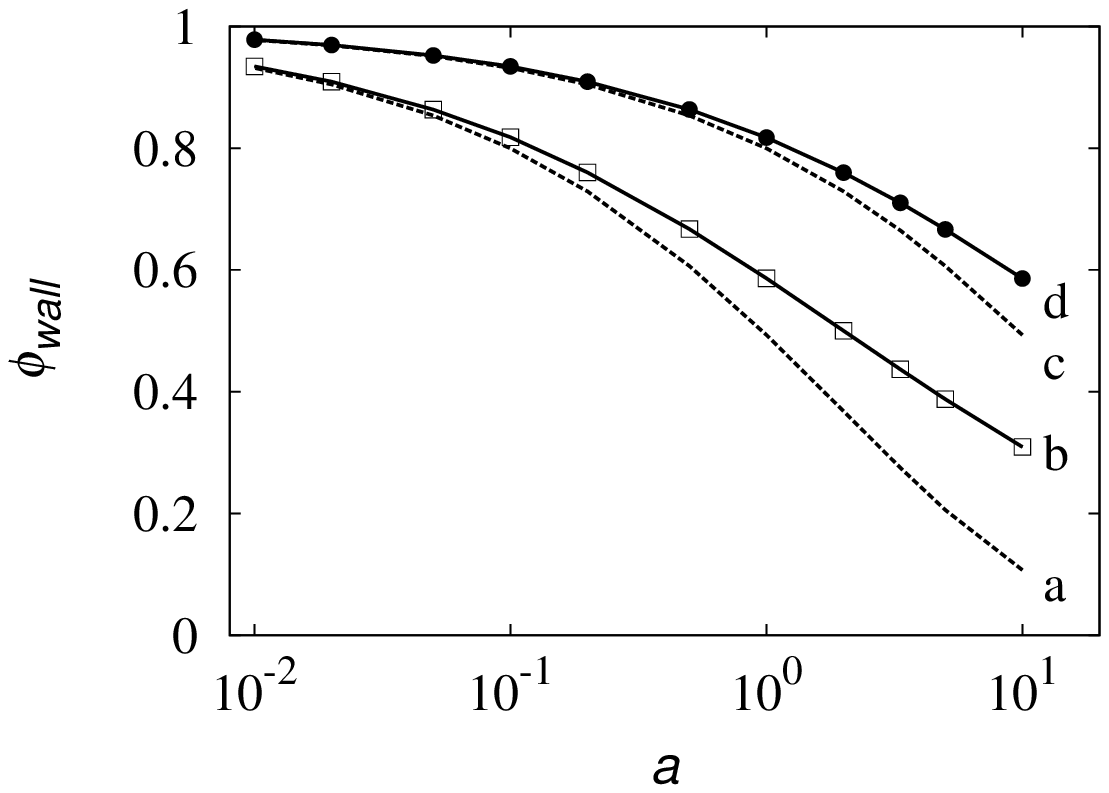}
\end{center}
\caption{$\phi_{\rm wall}$ vs the switching rate $a$ for
two values of $D$. Symbols ($\square$, $\bullet$) represent
simulation results, dashed lines the lower bound eq. (\ref{eq3_4}).
Lines (a) and (b) and symbols ($\square$): $D=1$,
lines (c) and (d) and symbols ($\bullet$): $D=10$.}
\label{Fig2}
\end{figure}
The estimate $\phi_0(a,D)$ approaches better the value $\phi_{\rm wall}$
at lower values of $a$ and higher values of $D$.

\section{Polarization model}
\label{sec_4}

As a second example, consider the one-dimensional model in the presence
of a deterministic biasing
field,
\begin{equation}
d x(t) = v(x(t)) \, dt + b \, (-1)^{\chi(t)} \, dt \, ,
\label{eq4_1}
\end{equation}
defined for $x \in [0,1]$,  equipped with impermeability
conditions at $x=0,1$. The velocity vield $v(x)$ is given by
\begin{equation}
v(x)= - v_m \, \sin \left ( \frac{3 \pi x}{2} \right ) \, ,
\label{eq4_2}
\end{equation} 
and $v_m = \gamma \, b$, with $\gamma \in (0,1)$, and $a$ and $b$ are
such that $D=1$. This case has been analyzed elsewhere \cite{brasiello}, 
as a model
for describing polarization effects within the Poisson-Kac paradigm,
arising as a consequence of a nonvanishing value of the deterministic
velocity field at one of the boundary points ($x=1$), determining
the formation of a polarization layer in the neighbourhood
of $x=1$.

In this article, we use this model to investigate better the 
conditions to be imposed on  the stochastic dynamics  (\ref{eq4_1})
in the presence of a nonvanishing deterministic bias at the boundary.

First, consider the partial waves associated with eq. (\ref{eq4_1}),
i.e. its Eulerian description.
The overall probability density flux $J_{\rm tot}(x,t)$ is the
sum of the convective (deterministic) and stochastic
contribution,
\begin{equation}
J_{\rm tot}(x,t) = v(x) \, \left [ p^+(x,t)+ p^-(x,t) \right ]
+ b \, \left [ p^+(x,t)- p^-(x,t) \right ] \; .
\label{eq4_3}
\end{equation}
Since $v(0)=0$, the zero-flux condition at $x=0$ simply becomes
\begin{equation}
p^+(0,t) = p^-(0,t) \, . 
\label{eq4_4}
\end{equation}
This is the total reflection condition.  At $x=1$, $v(1)=\gamma \, b$, and
the zero-flux boundary condition becomes in terms of partial waves
\begin{equation}
p^-(1,t)= \frac{1+\gamma}{1-\gamma} \, p^+(1,t) \, .
\label{eq4_5}
\end{equation}
Eq. (\ref{eq4_5}) is valid for $\gamma <1$. The extension to $\gamma>1$
is discussed in \cite{brasiello}, and therefore it is not repeated here.

For the stochastic differential equation (\ref{eq4_1})
the conditions at the boundary are simply reflection
conditions, corresponding to the mirror simmetric reflection 
of the particle position at $x=0,1$, and to the
switching of $(-1)^{\chi(t)} \mapsto -(-1)^{\chi(t)}$ of
the Poissonian perturbation.

In order to check the validity of these conditions, we compare the
stationary solutions $p_*^{\pm}(x)$ of the associated partial
wave model (\ref{eq2_9}) with $v(x)$ given by eq. (\ref{eq4_2}),
in the presence of the boundary conditions (\ref{eq4_4})-(\ref{eq4_5}),
and the overall stationary probability density function 
$p_*(x)=p_*^+(x)+p_*^-(x)$, with the corresponding stationary distributions
obtained by integrating the stochastic equation
of motion (\ref{eq4_1}) with an integration step $h_t=10^{-4}$
 for  an ensemble of $N=10^7$ particle in which the above mentioned
simple reflection conditions have been used at the boundaries.
This comparison is depicted in figure \ref{Fig3} and is 
fully satisfactory. 
\begin{figure}[h!]
\begin{center}
\epsfxsize=10cm
\epsffile{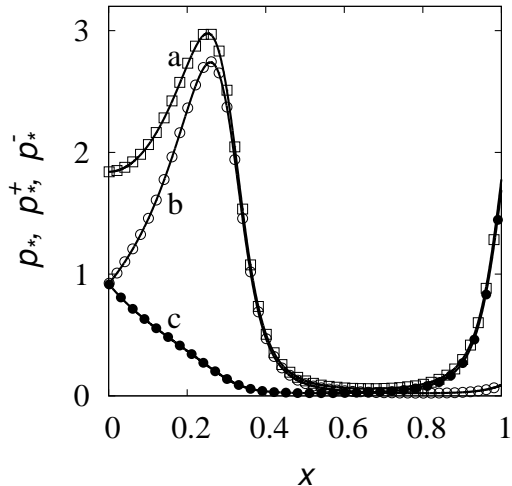} 
\end{center}
\caption{Stationary probability distribution 
$p_*(x)$ (lines (a) and ($\square$)),
and stationary partial waves $p_*^+(x)$ 
lines (b) and ($\circ$)), $p_*^-(x)$ (lines (a) and ($\bullet$))
for the polarization model (\ref{eq4_1}) at $D=1$, $a=5$, $\gamma=0.9$.
Solid lines represent the results of the integration of the partial wave equations,
symbols the results of stochastic simulations of eq. (\ref{eq4_1}).}
\label{Fig3}
\end{figure}

In point of fact,
a more stringent test on the influence of the boundary
conditions (especially at $x=1$) is based on the analysis
of the ratio $p_*^-(x)/p_*^+(x)$, that, because
of eqs. (\ref{eq4_4}) and (\ref{eq4_5}) should
attains the values $1$ at $x=0$, and $(1+\gamma)/(1-\gamma)$ at
$x=1$. Figure \ref{Fig4} shows the estimate of this ratio
obtained from the stochastic  evolution of the particle ensemble
compared with the corresponding quantity obtained 
from the partial wave model at different values of $\gamma$.

\begin{figure}[h!]
\begin{center}
\epsfxsize=10cm
\epsffile{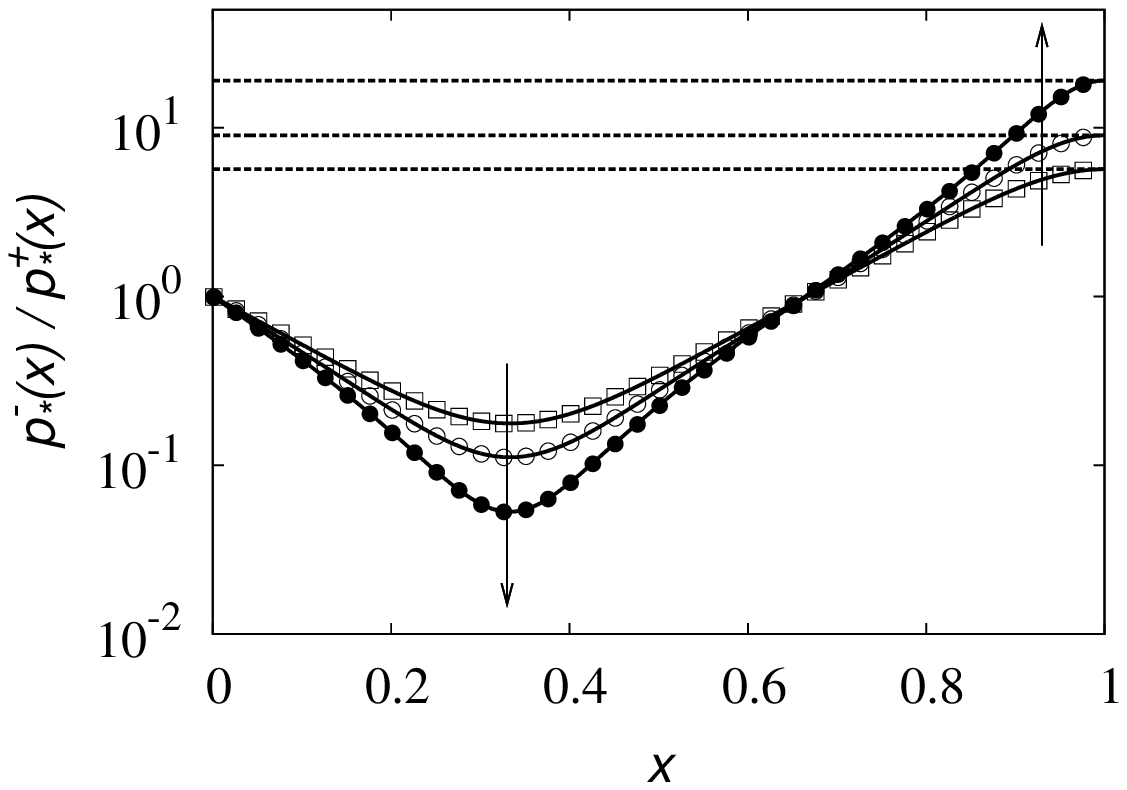} 
\end{center}
\caption{ Ratio  $p_*^-(x)/p_*^+(x)$  under
stationary conditions at $D=1$, $a=5$.
 Solid lines represent the results of the integration of the partial wave
 equations,
symbols the results of stochastic simulations of eq. (\ref{eq4_1}).
Arrows indicate increasing values of $\gamma=0.7,\,0.8,\, 0.9$.
Dashed horizontal lines corresponds to
the limit value at $x=1$, $(1+\gamma)/(1-\gamma)$.}
\label{Fig4}
\end{figure}

The dashed horizontal lines correspond to the limit values
 $(1+\gamma)/(1-\gamma)$ at $x=1$ for the values of $\gamma$ considered.
Also in this more severe test, the stochastic model equipped with
the simple reflecting boundary conditions provide excellent
quantitative agreement with the Eulerian results.

The switching time distributions $g(\tau)$ obtained from
stochastic simulations is depicted in figure \ref{Fig5} for several
values of $\gamma$. Compared with the corresponding results obtained
for the simple Knudsen model described in Section \ref{sec_3}, there
are qualitative analogies and differences.
The analogies are: (i) the compactness of the support of
$g(\tau)$, and (ii) the  occurrence of an impusive contribution
centered at $\tau_{\rm max}$. The 
differences are: (i) a significant modification in the
low-$\tau$ region induced by the polarization effects of the velocity field 
near the boundaries, and (ii) the exponential ``backbone'' of
this distribution, that apart for impulsive and nearly impulsive
contribution possesses an exponential decay equal to $a$, i.e., equal
to the bare Poissonian switching rate (compare the decay of the 
Poissonian statistics, line 
(e) with mean local decay of the other lines (a) to (d)).

\begin{figure}[h!]
\begin{center}
\epsfxsize=11cm
\epsffile{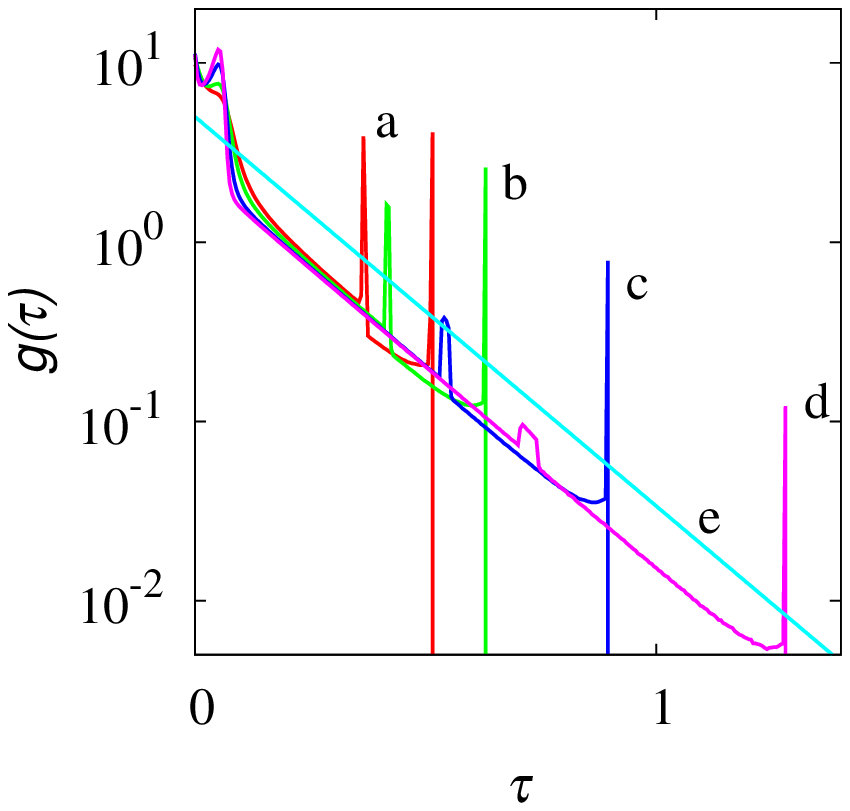}
\end{center}
\caption{Switching time probability density function $g(\tau)$
vs $\tau$ for the polarization model (\ref{eq4_1})-(\ref{eq4_2}) at $a=5$.
Line (a): $\gamma=0.7$, line (b): $\gamma=0.8$, line (c): $\gamma=0.9$,
line (d) $\gamma=0.95$. Line (e) represents the pure Poissonian
distribution $g_{P}(\tau)=ae^{-a \, \tau}$.}
\label{Fig5}
\end{figure}

\section{Kac diffusion on fractals}
\label{sec_5}

As a final example, consider diffusion in complex geometries
such as in two-dimensional connected
sets possessing fractal (Hausdorff) dimension $ 1< d_f<2$.
It is known that Brownian diffusion in fractal media possesses
anomalous properties \cite{havlin,havlin1}, typically expressed
by the power-law scaling of the mean square displacement
$\langle r^2(t) \rangle$ of
Brownian particles
\begin{equation}
\langle r^2(t) \rangle \sim t^{2/d_w} \, ,
\label{eq5_1}
\end{equation}
where $d_w >2$ is the walk dimension ($d_w=2$ in Euclidean media,
giving rise to the classical Einsteinian linear relation
of $\langle r^2(t) \rangle \sim t$).

Most of the existing numerical results on diffusion in fractal
media refer to lattice simulations, where Brownian particles move
in a random and uncorrelated way from a site of the fractal set
to one of its neighbouring sites belonging to the set itself.

In order to give concrete examples, consider the two fractal
structures depicted in figure \ref{Fig5a}: the Sierpinski carpet,
and a loopless deterministic fractal, henceforth referred to as the
deterministic cross fractal. Their fractal dimensions are
$d_f=\log 8/\log 3 \simeq 1.893$ (for the Sierpinski carpet), and
$d_f=\log 5/\log 3 \simeq 1.465$ for the deterministic cross fractal.
Anomalous  diffusion properties on the Sierpinski carpets
have been addressed in \cite{dasgupta,dasgupta1}.

\begin{figure}[h!]
\begin{center}
\hspace{-2.0cm}
\epsfxsize=8cm
\epsffile{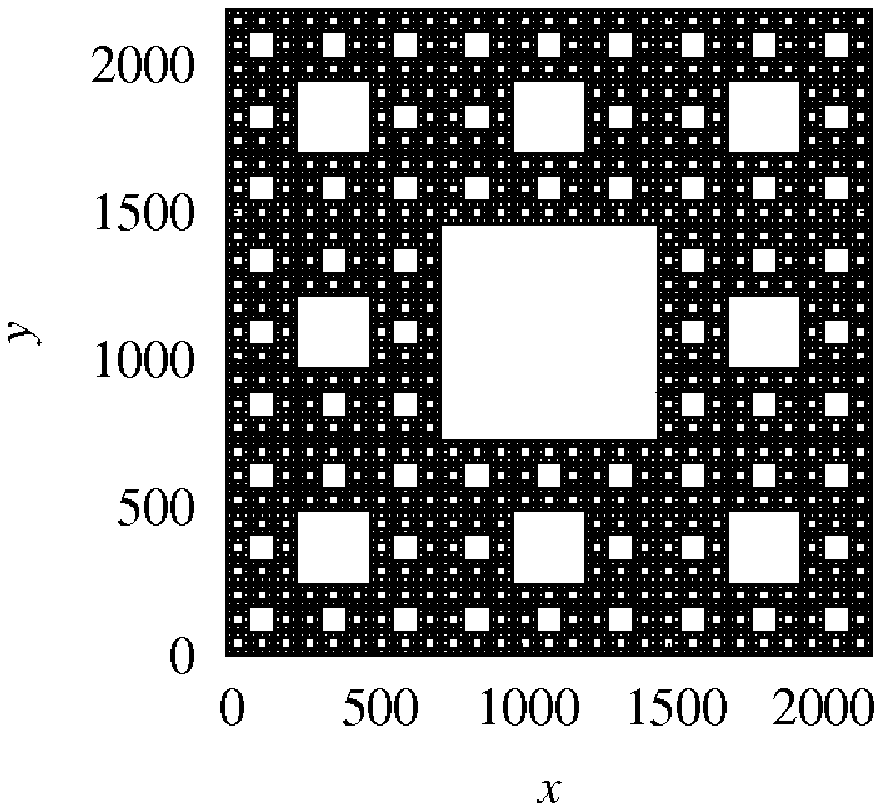} \hspace{-2.0cm} {\large (a)} 
\epsfxsize=8cm
\epsffile{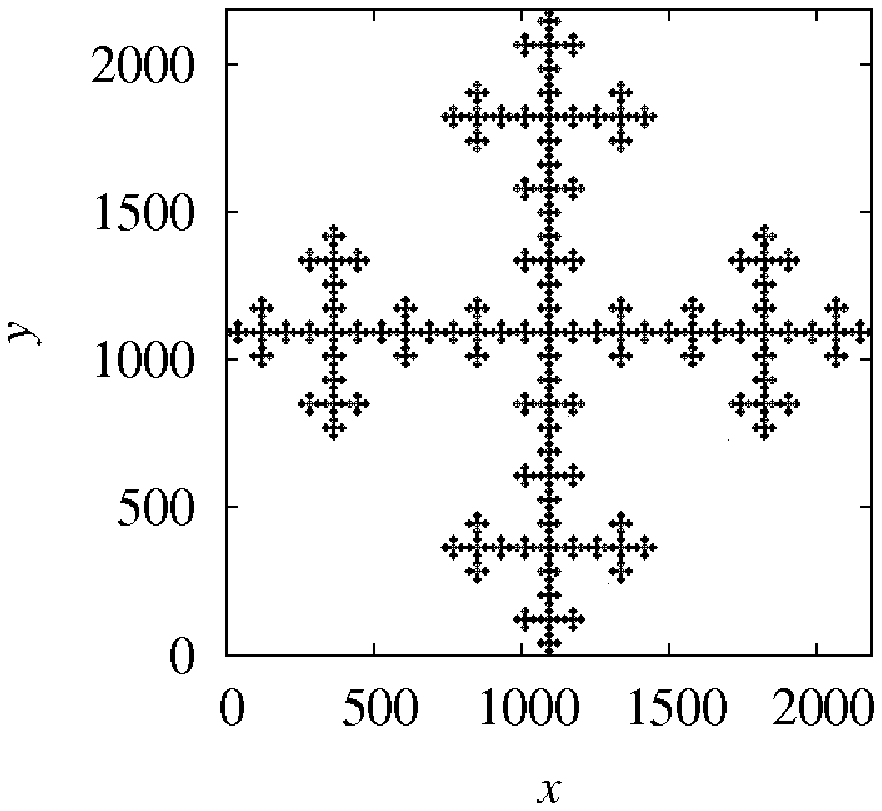} \hspace{-2.0cm} {\large (b)} 
\hspace{-2.0cm}
\end{center}
\caption{Panel (a): Sierpinski carpet. Panel (b): deterministic cross fractal.}
\label{Fig5a}
\end{figure}

Figure \ref{Fig6} depicts the behavior of
the mean square displacement of classical Brownian particles
obtained from discrete lattice simulations. The simulations
refer to a prefractal approximation of the structures at the $7$-th 
iteration of the costruction process, 
that corresponds to lattices possessing $3^7=2187$ lattice size per
Cartesian coordinate. In point of fact, figure \ref{Fig5a} depicts just
 these prefractal approximations, as can be noticed by
the labels of $x$ and $y$ coordinates.

\begin{figure}[h!]
\begin{center}
\epsfxsize=11cm
\epsffile{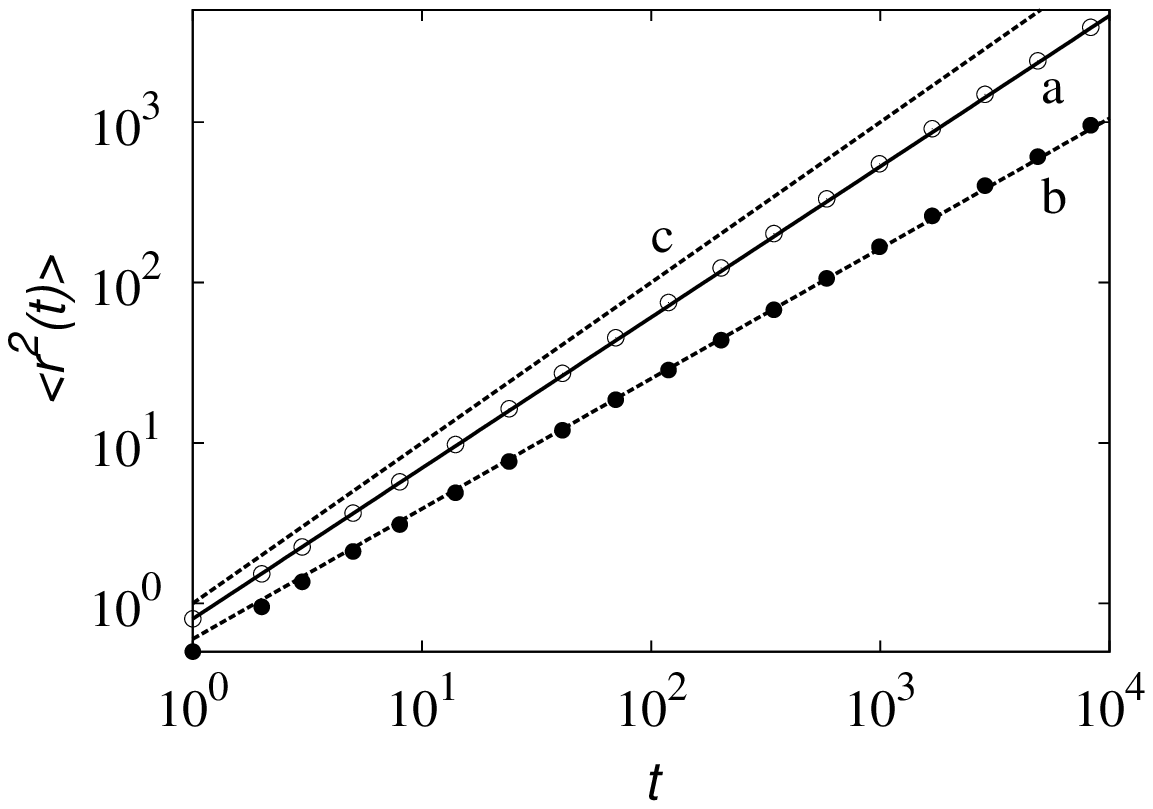}
\end{center}
\caption{Mean square displacement $\langle r^2(t) \rangle$ vs $t$
for classical Brownian
 lattice diffusion on the fractal structures considered.
Symbols are the results on lattice random  simulations: ($\circ$):
Sierpinski carpet, ($\bullet$): deterministic cross fractal.
Lines (a) and (b) represent the scaling $\langle r^2(t) \rangle \sim t^{2/d_w}$,
with $2/d_w=0.94$ (line (a)), and $2/d_w=2/(d_f+1)$, $d_f=\log 5/\log 3$ (line (b)). 
Line (c) represents $\langle r^2(t) \rangle  \sim t$.}
\label{Fig6}
\end{figure}

From the scaling of the mean square displacement (line a)
we obtain the numerical value $2/d_w=0.94$
for the Sierpinski carpet (no analytic results are available for the
Sierpinski carpet, which is an infinitely ramified structure not
easily amenable to exact real-space renormalization).
In the case of the deterministic cross fractal, the scaling theory
of loopless fractals, based on the concept of chemical distance, predicts
a simple relation between $d_w$ and $d_f$, namely \cite{havlin,havlin1}
\begin{equation} 
d_w = d_f + 1 \, ,
\label{eq5_2}
\end{equation}
and the numerical results perfectly agree with the scaling
$\langle r^2(t) \rangle \sim t^{2/(d_f+1)}$ (line b).

Next, consider Poisson-Kac diffusion in these sets. The mathematical
setting of the problem has been developed in paragraph
\ref{sec_2_1}, both as it regards the stochastic equations
of motion and their statistical (Eulerian) description
in terms of partial waves.
In the numerical simulations of eq. (\ref{eq2_16}) we
consider  that the unit site of the prefractal approximation possesses
a unit linear length. Consequently, the unit square is the
building block of the fractal structure, that at iteration
$n=7$ extends over a length of $L=3^7=2187$.
Using eq. (\ref{eq2_16}), and applying total reflection conditions
whenever a particle hits the boundary (which implies also the switching
of $(-1)^{\chi_1(t)}$ or $(-1)^{\chi_2(t)}$ depending on the
which boundary is involved),
 it is possible to perform an off-lattice (continuous)
simulation of the Poisson-Kac diffusion process. Let  $D=1$ be  the
nondimensional diffusion coefficient.
Figure \ref{Fig7} panels (a) to (c) depict a portion of an orbit
of a Poisson-Kac particle diffusing inside the Sierpinski carpet at
different values of $a$. 
An analogous picture for the
deterministic cross fractal is shown in  figure \ref{Fig8}.

\begin{figure}[!]
\begin{center}
\hspace{-2.0cm}
\epsfxsize=10cm
\epsffile{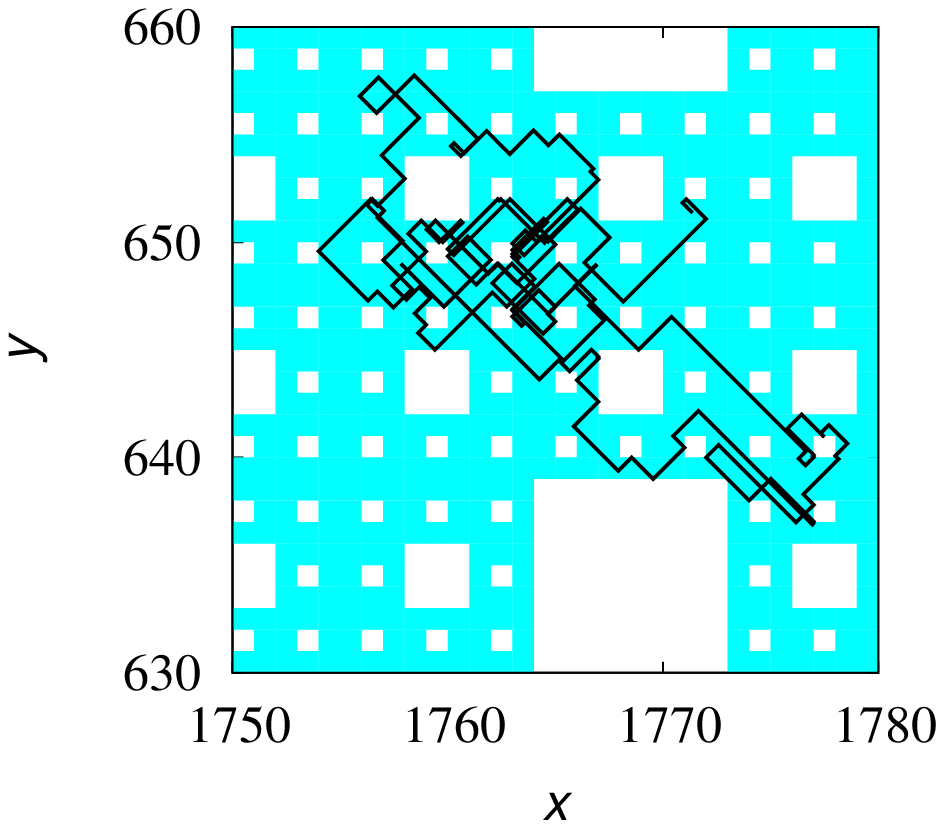} \hspace{-2.0cm} {\large (a)} \\
\hspace{-2.0cm}
\epsfxsize=10cm
\epsffile{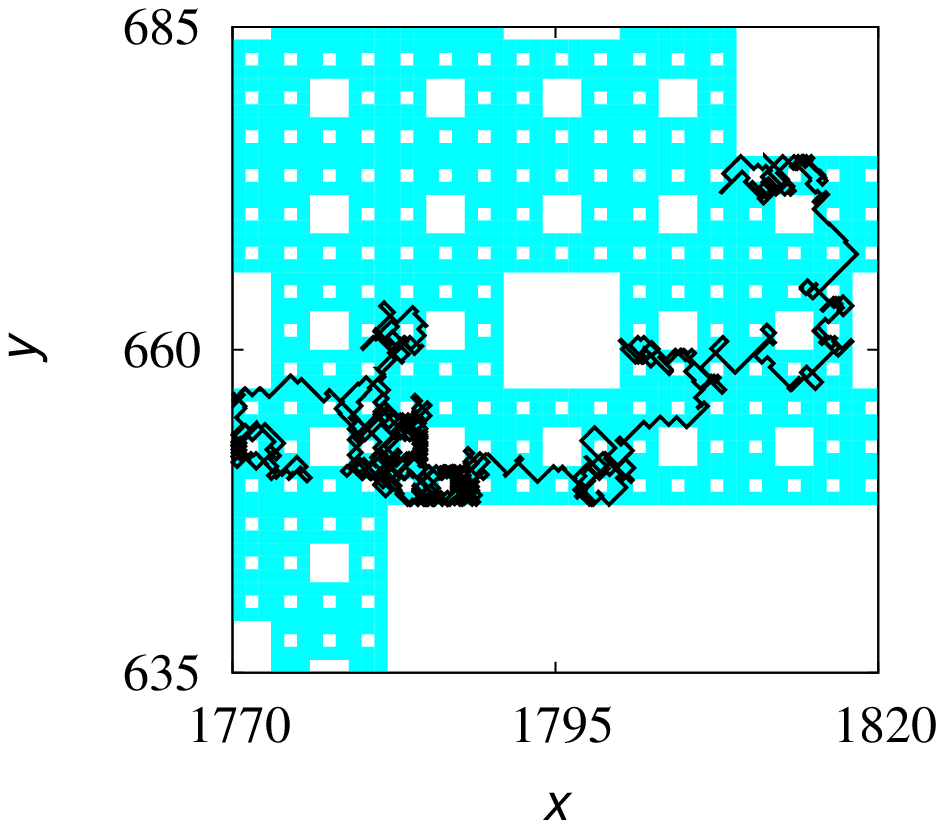} \hspace{-2.0cm} {\large (b)} \\
\hspace{-2.0cm}
\epsfxsize=10cm
\epsffile{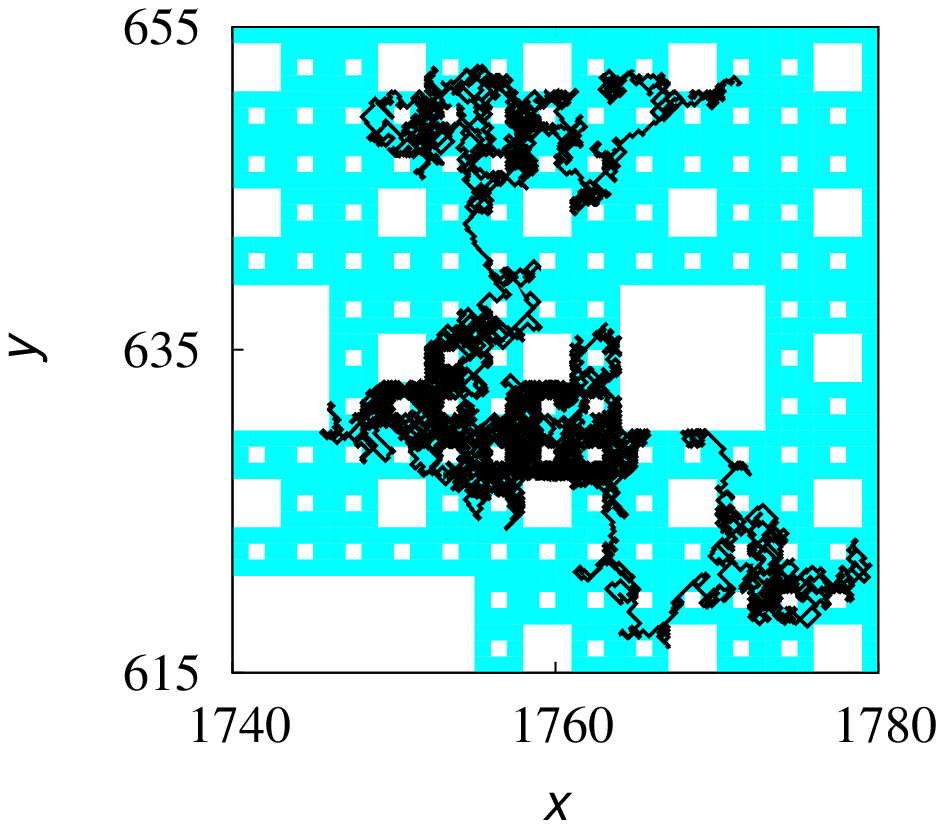} \hspace{-2.0cm} {\large (c)}
\end{center}
\caption{Realizations of  diffusive Kac orbits on the Sierpinski
carpet at $D=1$. Panel (a): $a=0.1$, panel (b): $a=1$, panel (c) $a=10$.}
\label{Fig7}
\end{figure}

\newpage
\begin{figure}[h!]
\begin{center}
\epsfxsize=11cm
\epsffile{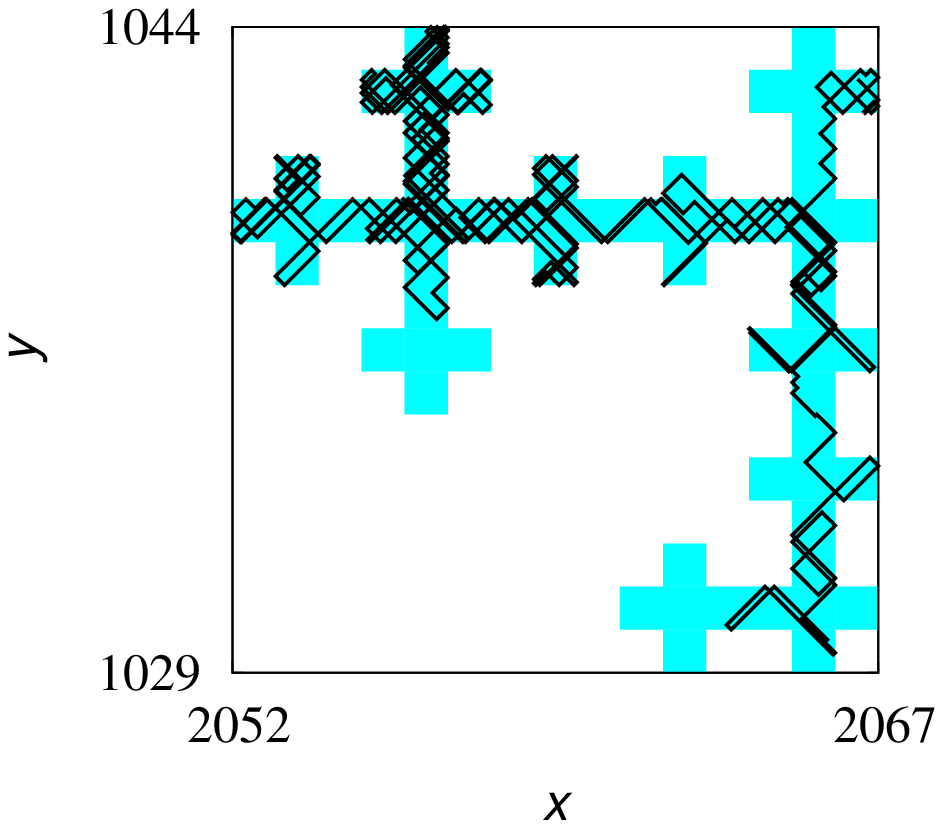}
\end{center}
\caption{Realization of  a diffusive Kac orbit on the 
deterministic cross fractal at $D=1$ and  $a=0.1$.}
\label{Fig8}
\end{figure}

At small  values of $a$, figure \ref{Fig7}  panel (b) and figure \ref{Fig8}, 
(keeping $D$ fixed and consequently,
 varying   $b$ according to the 
 Kac condition eq. (\ref{eq2_13})), the orbits 
are significantly different from that of classical Brownian motion, as
the switching of the Poissonian processes is controlled by 
collisions with the boundary of the fractal  structure.
As $a$ increases (see panel (b) and (c) in figure \ref{Fig7}),
particle orbits resemble more closely that of classical
Brownian particles.
Figure \ref{Fig9} depict the mean square displacement of
Poisson-Kac particles in the two fractal  structures considered
at several values 
of the switching rate $a$. The mean square displacement
 $\langle r^2(t) \rangle$
possesses two asymptotic scalings: at short time scales the 
mean square displacement is ballistic, and this is to be
expected from the wave nature of the stochastic process, as
for $t <<\min\{1/b, 1/a\}$ particles move in straight lines.
In the long-term limit, the mean square displacement possesses
exactly the same anomalous behavior characterizing classical
Brownian particles.

\begin{figure}[h!]
\begin{center}
\hspace{-2.0cm}
\epsfxsize=10cm
\epsffile{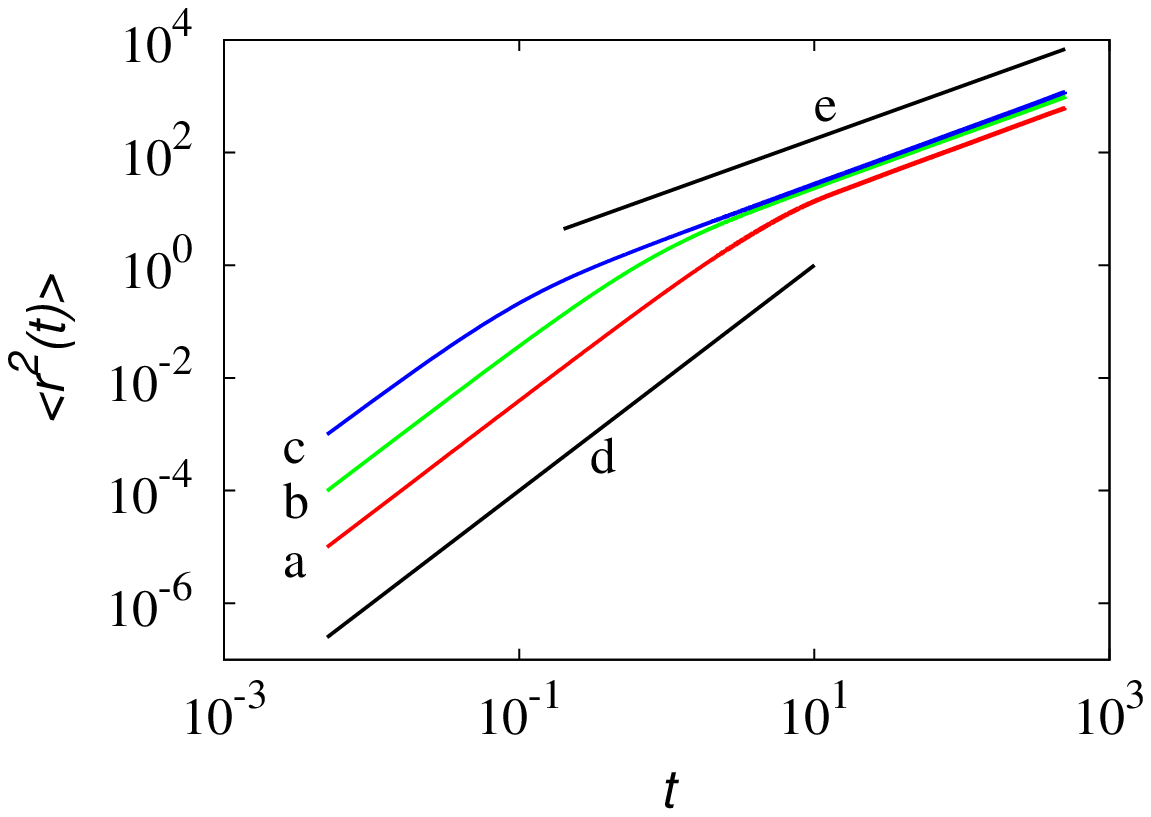} \hspace{-2.0cm} {\large (a)} \\
\hspace{-2.0cm}
\epsfxsize=10cm
\epsffile{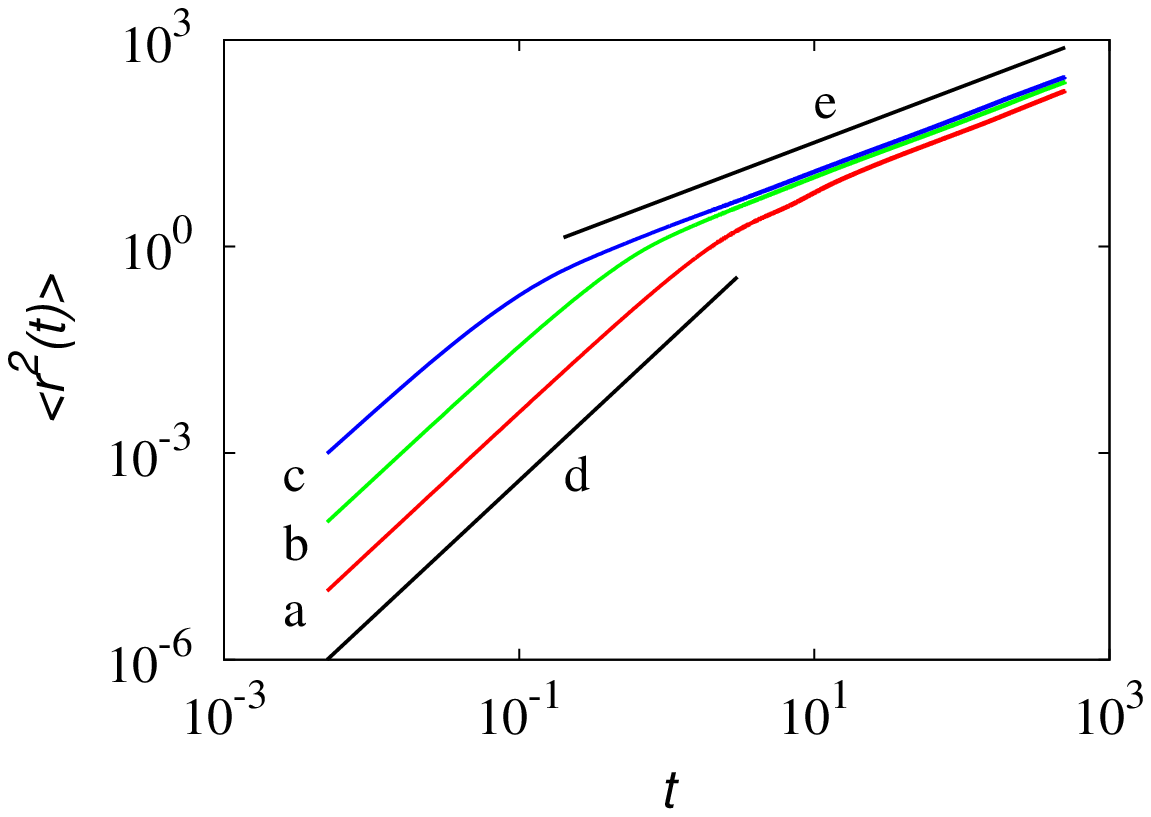} \hspace{-2.0cm} {\large (b)} 
\end{center}
\caption{Mean square displacement $\langle r^2(t) \rangle$ vs time $t$ for
Kac diffusion on Sierpinski carpet (panel a) and on the
deterministic cross fractal (panel b) at $D=1$. Lines (a) to (c) refer to 
different values of $a=0.1$ (lines a), $a=1$ (lines b), $a=10$ (lines c).
Lines (d) correspond to the scaling $\langle r^2(t) \rangle \sim t^2$.
Lines (e) to the  anomalous diffusion scaling $\langle r^2(t) \rangle \sim t^{2/d_w}$, where $2/d_w=0.94$ (panel a), and $2/d_w=0.8114$ (panel b).}
\label{Fig9}
\end{figure}

Anomalous diffusion effects, i.e. the occurrence of  a value of $d_w >2$,
are in the Poisson-Kac case the emerging feature  of the collisions with the
boundary of the fractal structure, and therefore
it is to be expected that the switching time statistics becomes significantly
modified. As the process is two-dimensional, we consider separately the
statistics $g_\alpha(\tau)$, $\alpha=1,2$ for the switchings of the
two Poissonian processes. The hit of the boundaries orthogonal to the 
$x$-axis determines a switching of $(-1)^{\chi_1(t)}$, and the
reflection onto a boundary orthogonal to the $y$-axis determines a switching
of  $(-1)^{\chi_2(t)}$.

\begin{figure}[h!]
\begin{center}
\hspace{-2.0cm}
\epsfxsize=8cm
\epsffile{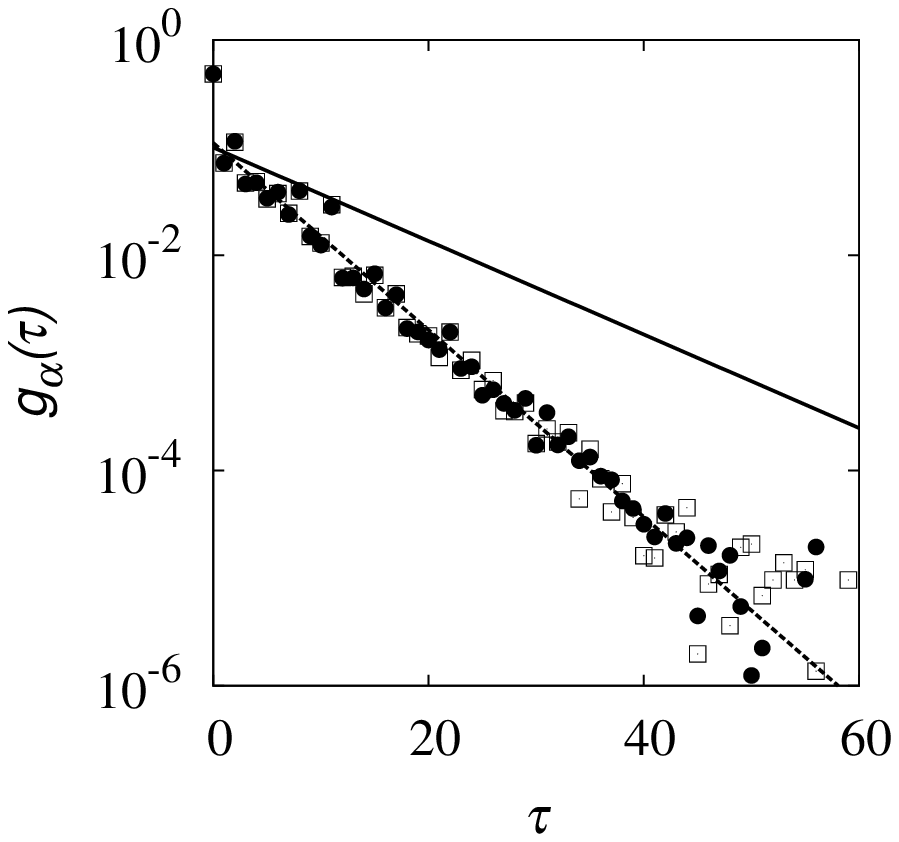} \hspace{-2.0cm} {\large (a)} 
\epsfxsize=8cm
\epsffile{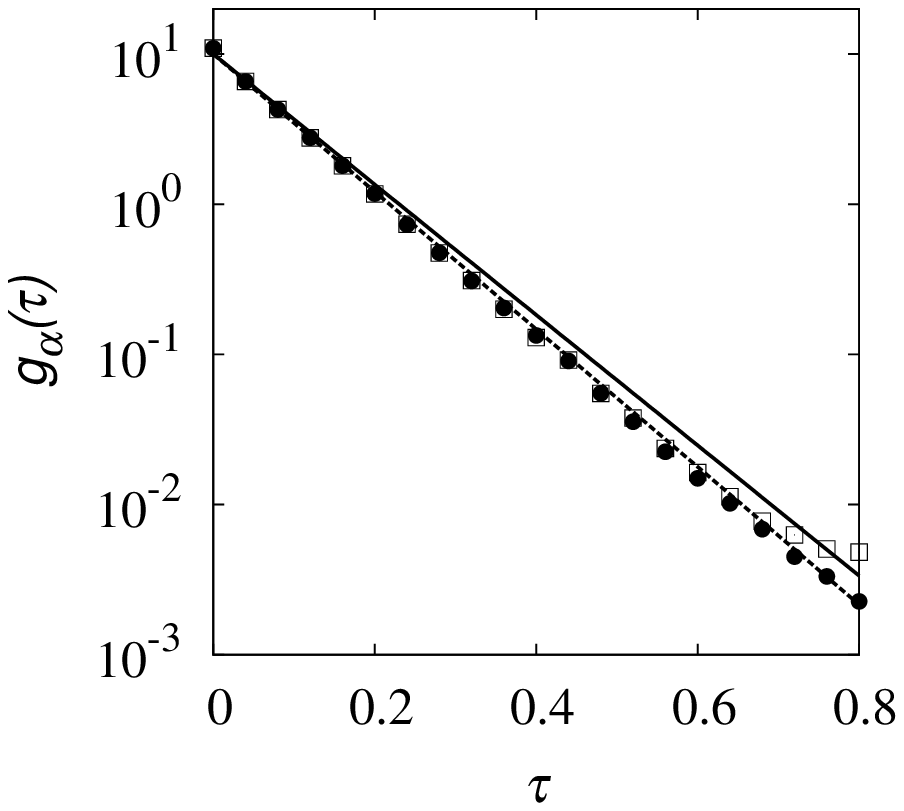} \hspace{-2.0cm} {\large (b)}  
\end{center}
\caption{Switching time probability density functions $g_\alpha(\tau)$, $\alpha=1,2$, vs $\tau$
for diffusion on the Sierpinski carpet at $D=1$. Panel (a): $a=0.1$,
panel (b): $a=10$.  Symbols ($\square$) refer to $g_1(\tau)$,
symbols ($\bullet$) to $g_2(\tau)$. The solid line is the bare Poisson
statistics $g_\alpha(\tau)=a \, e^{-a t}$, the dashed line is the
exponential fit of the data.}
\label{Fig10}
\end{figure}
Figure \ref{Fig10} and \ref{Fig11} depict the behavior of
$g_\alpha(\tau)$ $\alpha=1,2$ obtained from stochastic simulations.
The two density functions $g_1(\tau)$, $g_2(\tau)$ are equal and differ
from the pure Poissonian statistics $g_P(\tau)$. Nevertheless, their
behavior is well approximated by an exponential decay with an effective
decay rate $a_e$ different from $a$.
\begin{figure}[h!]
\begin{center}
\epsfxsize=11cm
\epsffile{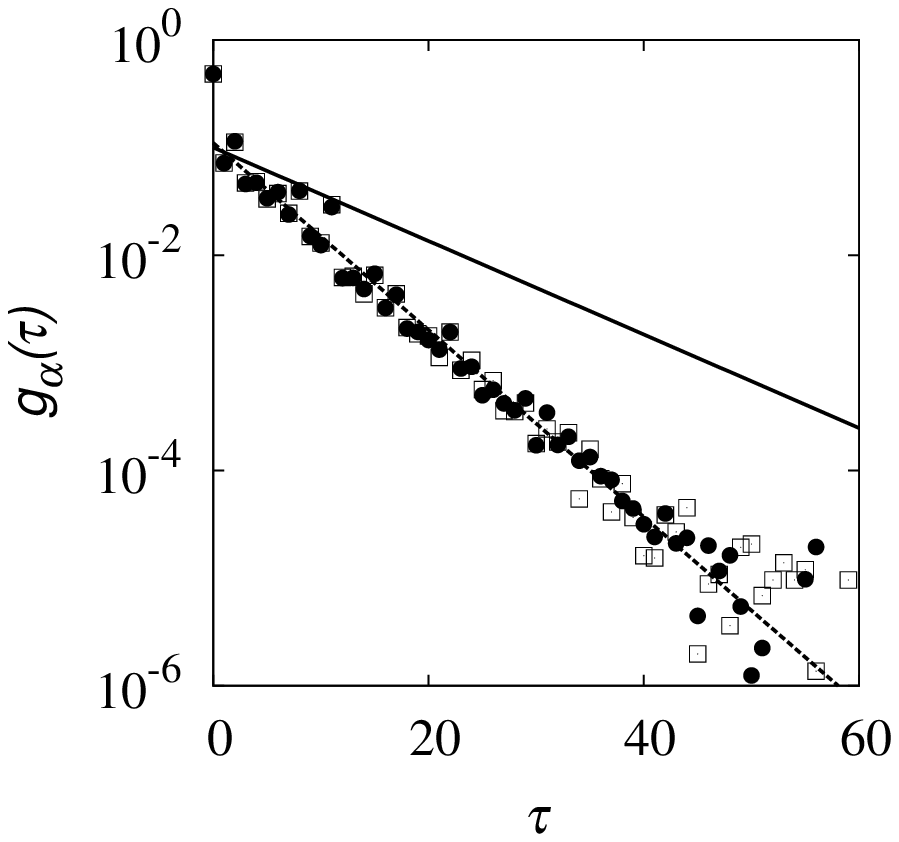} 
\end{center}
\caption{Switching time probability density functions $g_\alpha(\tau)$, $\alpha=1,2$, vs $\tau$
for diffusion on the deterministic cross fractal at $D=1$, $a=1$. 
Symbols ($\square$) refer to $g_1(\tau)$,
symbols ($\bullet$) to $g_2(\tau)$. The solid line is the bare Poisson
statistics $g_\alpha(\tau)=a \, e^{-a t}$, the dashed line is the
exponential fit of the data.}
\label{Fig11}
\end{figure}

A coarse information of the role of the boundary collisions can
be obtained by considering the ratio $\phi_{\rm coll}$ of the overall
number of collisions with the walls to the number of  switchings in the velocity
determined by the Poissonian statistics.
The behavior of $\phi_{\rm coll}$ vs $a$ is depicted
in figure \ref{Fig12} at $D=1$. It can be observed that this
ratio follows a power law behavior $\phi_{\rm coll}(a) \sim a^{-\zeta}$,
where the exponent $\zeta$  attains the value (from the best fit of simulation
data): $\zeta=0.72$ for the Sierpinski carpet, $\zeta=0.79$ for the
deterministic cross fractal. This power law scaling is nor related in a
simple way to the characteristic geometric ($d_f$) and dynamic ($d_w$)
exponents of the medium. We leave a more careful theoretical analysis
on the properties of this scaling law, and of its framing
within the theory of transport in fractal media to future investigations.
\begin{figure}[h!]
\begin{center}
\epsfxsize=11cm
\epsffile{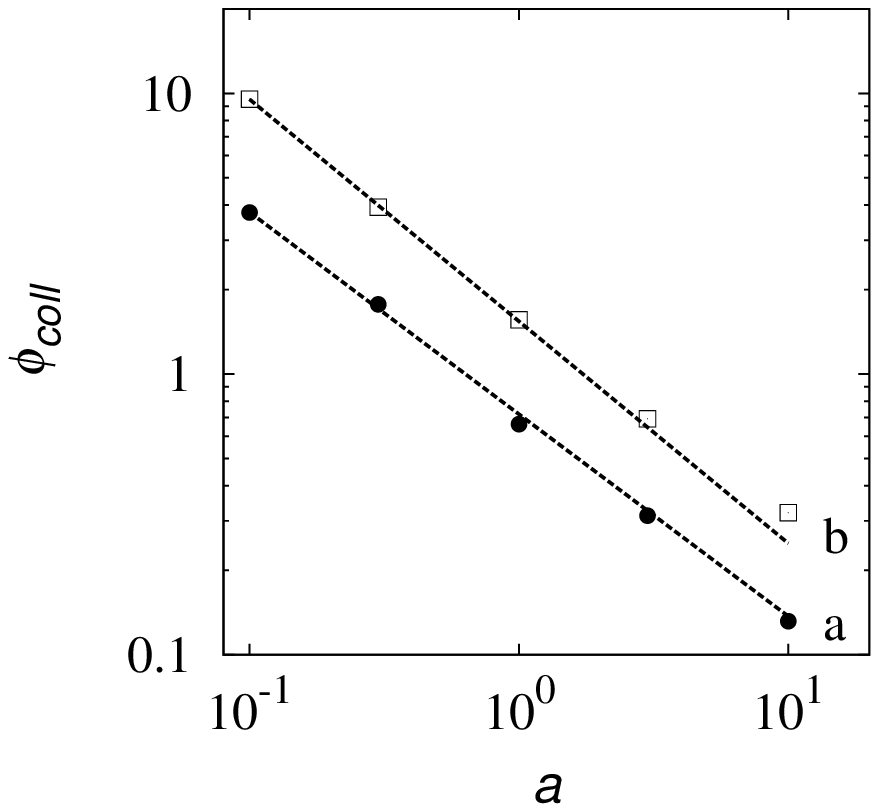} 
\end{center}
\caption{$\phi_{\rm coll}$ vs $a$ for Kac diffusion on fractals at $D=1$.
Solid lines represent the scaling $\phi_{\rm coll}(a) \sim a^{-\zeta}$.
Line (a) and ($\bullet$) refer to the Sierpinski carpet ($\zeta=0.72$),
line (b) and ($\square$) to the deterministic cross fractal ($\zeta=0.79$).}
\label{Fig12}
\end{figure}

Let us analyze in more detail the trajectories of Poisson-Kac
particles on fractals. By definition, the trajectories of the
realizations  of Poisson-Kac processes are with probability $1$ almost
eveywhere smooth curves. Discontinuities in the derivatives arise
as a consequence of the ``internal'' Poissonian switching occurring
 in average
at a characteristic time $\tau_c=1/a$, and as a
result of the ``external'' reflections with the
boundaries. Since the unit building block of the fractal structure
possesses unitary  characteristic length, the
average time between collisions with the boundary is $\tau_w=
1/b= 1/\sqrt{2 D a}$. 

For time scales much smaller than the minimum between $\tau_c$ and $\tau_w$,
the trajectories are simply straight lines. Figure \ref{Fig13}
depicts two examples of particle trajectory  (actually its $x$-component)
at two different values of $a$. The higher $a$ is, the smaller is the
characteristic timescale at which the realization of Poisson-Kac 
processes attain asymptotic fractal properties.

\begin{figure}[h!]
\begin{center}
\epsfxsize=11cm
\epsffile{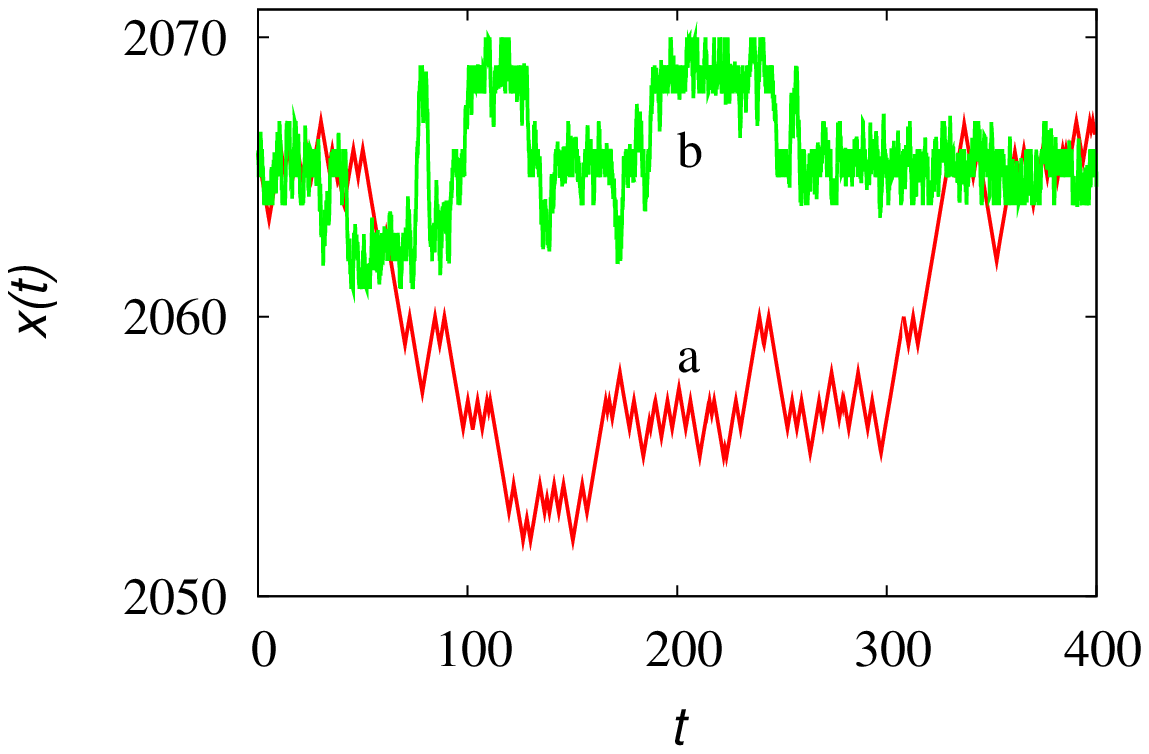}
\end{center}
\caption{$x$-component $x(t)$ of a diffusive Kac trajectory vs $t$
on the deterministic cross fractal at $D=1$. Line (a) refers to
$a=0.1$, line (b) to $a=10$.}
\label{Fig13}
\end{figure}

The role of wall collisions that superimpose to  the intrinsic 
Poissonian switchings is therefore to induce
in the long-term behavior of particle trajectories a fractal behavior
characterized by a fractal dimension different from $3/2$, that corresponds
to the classical Brownian motion in Euclidean media.

Let $d_T$ be the fractal dimension of Poisson-Kac trajectories
in the long-term regime, and $L(\Delta t;t_{\rm max})$  the length
of a finite portion of the trajectory in the interval $t \in [0,t_{\rm max})$
estimated using a temporal yardstick $\Delta t$. Henceforth,
for notational simplicity, we drop the reference to $t_{\rm max}$ and write
simply $L(\Delta t)$.

It follows from the above qualitative analysis that
\begin{equation}
L(\Delta t) \sim
\left \{
\begin{array}{lll}
\mbox{constant} & & \Delta t < \Delta t^* \\
\Delta t^{1 - d_T} & & \Delta t \gg \Delta t^* \, ,
\end{array}
\right .
\label{eq5_3}
\end{equation}
where the crossover time $\Delta t^*$ depends on $a$. 
The fractal dimension $d_T$ is related to the Hurst exponent of the graph
of the trajectories by the relation \cite{tricot}
\begin{equation}
d_T = 2 - H \, ,
\label{eq5_4}
\end{equation}
and, in turn, it is related to the walk dimension $d_w$ by $H=1/d_w$, so that
\begin{equation}
d_T= 2 - \frac{1}{d_w} \, .
\label{eq5_5}
\end{equation}
Consequently, the length-yardstick analysis, typical of the  geometric
characterization of fractal curves \cite{tricot}, provides an alternative way
to estimate $d_w$ from a single realization of the process.

Figure \ref{Fig14} depicts the graphs of $L(\Delta t)$ vs $\Delta t$
for several Poisson-Kac processes on the Siepinski carpet and on the
deterministic cross fractal at different values of $a$.
The realization of the process has been obtained by integrating 
eq. (\ref{eq2_16}) with a time step $h_t= 2 \times 10^{-3}$ up to
$t_{\rm max}=4 \times 10^2$, starting from a point inside the
fractal structure chosen at random.
\begin{figure}[h!]
\begin{center}
\epsfxsize=11cm
\epsffile{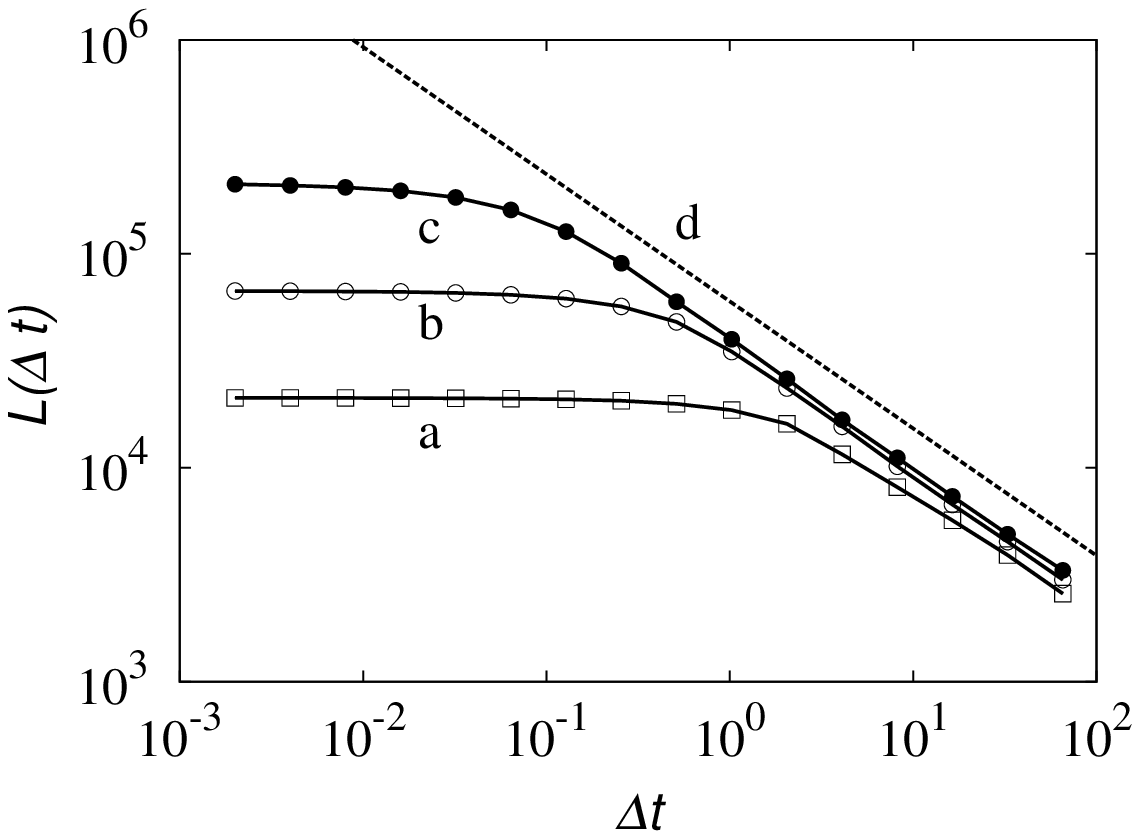} \hspace{-1cm} {\large (a)} \\
\epsfxsize=11cm
\epsffile{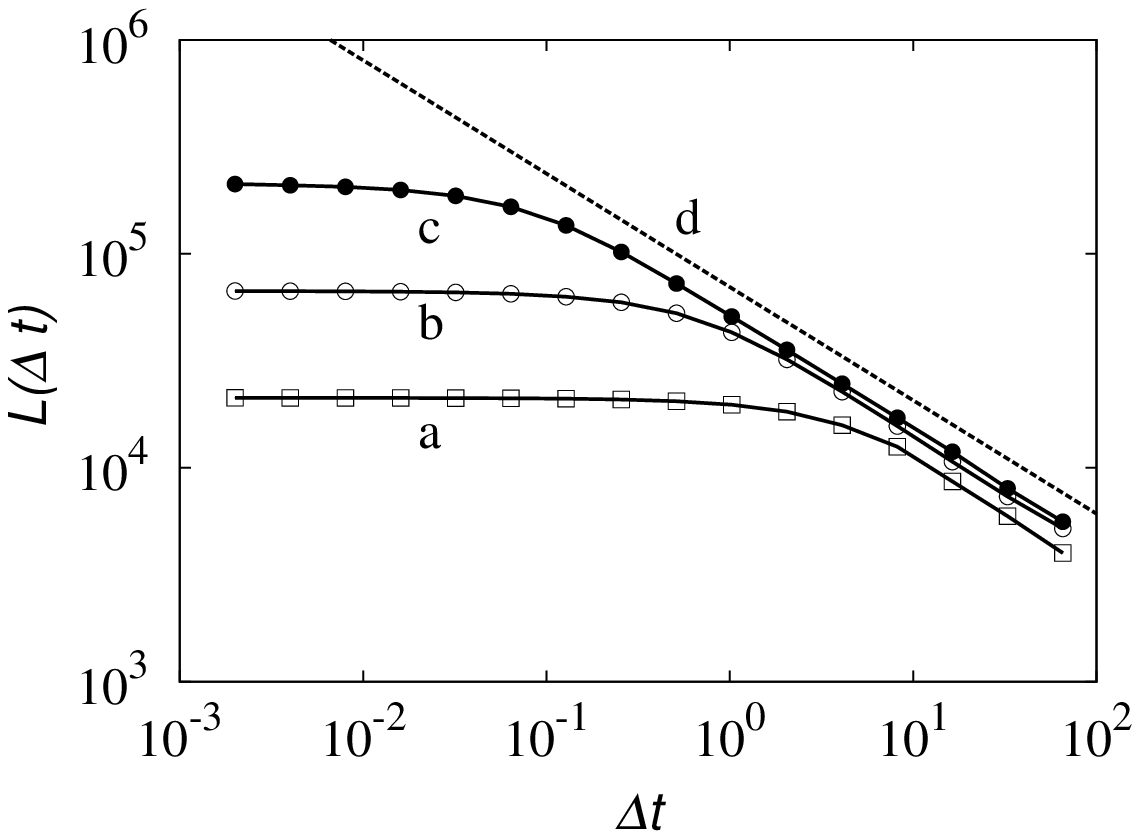} \hspace{-1cm} {\large (b)}
\end{center}
\caption{Scaling of the length $L(\Delta t)$ vs the temporal
 yardstick size $\Delta t$
on a diffusive Kac trajectory at $D=1$. Panel (a) refers to 
the deterministic cross fractal, panel (b) to the  Sierpinski
carpet. Lines (a) and ($\square$) correspond to $a=0.1$,
lines (b) and ($\circ$) to $a=1$, lines (c) and ($\bullet$) to
$a=10$. Lines (d) corresponds to the
theoretical prediction $L(\Delta t) \sim \Delta t^{1/d_w-1}$,
where $d_w=\log 5/\log 3+1$ in panel (a), and $d_w=2.4648$ in panel (b).}
\label{Fig14}
\end{figure}

The data depicted in figure \ref{Fig14} agree with the scaling (\ref{eq5_3}).
For small $\Delta t  < \Delta t^*$, where $\Delta t^*$ decreases with $a$,
$L(\Delta t)$ saturate, due to the rectifiable nature of the trajectories.
For $\Delta t > \Delta t^*$ the fractal scaling emerges as
a long-term property of the process. From eqs. (\ref{eq5_3})-(\ref{eq5_5})
it follows that
\begin{equation}
L(\Delta t) \sim \Delta t^{1/d_w -1} \, , \qquad \Delta t \gg \Delta t^* \, ,
\label{eq5_6}
\end{equation}
and the value of $d_w$ estimated from the long-term fractal
scaling of $L(\Delta t)$ using eq. (\ref{eq5_6}) perfectly
agrees quantitatively with the value of the walk dimension obtained
from the analysis of particle mean square displacement.

As expected, the fractal scaling appears more neatly (i.e.,
over a larger range of $\Delta t$) if one considers higher values of $a$
(curves (c) in figure \ref{Fig14}), since, for fixed $D$, the characteristic
stochastic velocity $b= \sqrt{2 \, D \, a}$ is higher and the
collisions with the fractal boundary, determining the occurrence
of the emergent fractal behavior of the trajectories,
 stabilize the process at smaller $\Delta t$.

\section{Concluding remarks}

\label{sec_6}

The statistics of Poisson-Kac processes in closed systems is profoundly
affected by  reflection conditions at the boundaries.
This effect influences either the statistics
of the switching times and the  emerging long-term
properties of the transport process. The study of Poisson-Kac
diffusion on fractals clearly shows that the
anomalies in diffusion occur as emerging properties
of a local, almost everywhere differentiable, stochastic
motion at microscales,  deriving from the complex collision
process occurring at the boundary of a fractal
set, and determining the transition from a ballistic motion
to anomalous Einstein scaling characterized by a walk dimension
greater than $2$.

The crossover in the mean square displacement has its
dynamic counterpart in the properties
of the trajectories deriving from the analysis of the scaling
of the length $L(\Delta t)$ vs the timescale $\Delta t$.
The trajectories of Poisson-Kac diffusive particles admit
a transition from a local smooth behavior at shorter timescales
to the emerging fractal properties characterized by a trajectory
fractal dimension $d_T=2-1/d_w$.

Two observations deserve further attention.
Poisson-Kac processes addressed in Section \ref{sec_5} 
are indeed useful tools to investigate, using off-lattice algorithms,
complex transport processes in fractal and disordered media, either
in the case of pure diffusive motion (as in the present work)
or including the effects of deterministic velocity fields
and potentials.
At relatively small values of $a$ and $b$, they represent 
the stochastic transport of a kind of a viscoelastic phase
with memory, and memory effects influence the short-term
behavior. Conversely the long-term, long-distance properties
of Poisson-Kac processes are those of classical Brownian motions.

However, Poisson-Kac
processes do not share with the classical Brownian motion
all the conceptual problems associated with an infinite propagation
velocity, and therefore they can be consistently applied also
to relativistic problems.

The second observation is purely technical. In order to highlight
the long-term fractality of Poisson-Kac trajectories emerging
from locally differentiable stochastic motion we have used
the length/time-interval scaling (figure \ref{Fig14}), out of
which the H\"{o}lder exponent $H$ of the stochastic trajectories can be
estimated. Since $H$ is related to the walk dimension $d_w$,
the scaling of $L(\Delta t)$
vs $\Delta t$ provides a  reliable, accurate and efficient way
to analyze anomalous transport properties in fractal media,
alternative to the more classical analysis of the scaling
of the mean square displacement $\langle r^2(t) \rangle$, just
using a single particle trajectory.

\bibliographystyle{model1-num-names}
\bibliography{<your-bib-database>}

\end{document}